\documentstyle[11pt]{article}
\newcommand{\blankline}{\vskip .3cm}
\newcommand{\f}{\begin{equation}}
\newcommand{\ff}{\end{equation}}
\begin{document}
\centerline{\LARGE The strong and weak holographic principles}
\blankline
\rm
\centerline{Lee Smolin${}^*$}
\blankline
\centerline{\it  Center for Gravitational Physics and Geometry}
\centerline{\it Department of Physics, The Pennsylvania State University}
\centerline{\it University Park, PA, USA 16802 \ \ and }
\centerline{\it The Blackett Laboratory,Imperial College of 
Science, Technology and Medicine }
\centerline{\it South Kensington, London SW7 2BZ, UK}
\vfill
\centerline{March 1, 2000}
\vfill
\centerline{ABSTRACT}
We review the  different proposals which have so far been made for the 
holographic principle and the related entropy bounds and classify
them into the strong, null and weak forms.  These are analyzed, 
with the aim of discovering which may hold at the level of the full 
quantum theory of gravity.  We find that only the weak forms, which 
constrain the information available to observers on boundaries, 
are implied by arguments using the generalized second law.  The strong forms, 
which go further and posit a bound on the entropy in spacelike regions bounded 
by surfaces, are found to suffer from serious problems, which give rise
to counterexamples already at the semiclassical level.   The null form, 
proposed by Fischler, Susskind, Bousso and others, in which the bound
is on the entropy of certain null surfaces, appears adequate at the 
level of a bound on the entropy of matter in a single background 
spacetime, but attempts to include the gravitational degrees of freedom 
encounter serious difficulties. Only the weak form seems capable of 
holding in the full quantum theory.  

The conclusion is that the holographic principle is not a relationship
between two independent sets of concepts: bulk theories and measures of
geometry vrs boundary theories and measures of information. Instead, it is the 
assertion that in a fundamental theory the first set of concepts  
must be completely reduced to the second.

\blankline

\blankline
${}^*$ smolin@phys.psu.edu
\eject
\tableofcontents
\eject

\section{Introduction}

The purpose of this paper is to examine the question:  {\it ``Exactly what
consequences do
the holographic principle\cite{thooft-holo,lenny-holo,louis-preholo}, and 
the related entropy bounds\cite{bek}, have for
the construction of the quantum theory of gravity?''}.  We will take 
it as given that a quantum theory of gravity should be both 
background independent and cosmological.  
This is because both background dependence and boundaries are almost
certainly artifacts of approximations which, while  
convenient for certain purposes, exclude significant aspects of the 
problem of constructing a theoretical framework which includes and 
extends the principles of both relativity and quantum 
theory\cite{morebi}.  

The question is not easy because most of the results which are so far 
known to bear on it are concerned with semiclassical approximations or
weak coupling limits.  Many are also limited to situations with 
boundaries, either asymptotic or finite.   It is, of course, always possible
that the holographic principle is only a characterization of the
semiclassical theory, perhaps because it is no more than a 
re-expression of the 
generalized second law of thermodynamics.  On the other hand, given 
that the entropy bounds involve inverse powers of $\hbar G$ it is very 
possible that they are deep clues to the structure of the fundamental 
theory, and that some version of the holographic principle may even
turn out to be a fundamental principle of the quantum theory of 
gravity.  If so it will be the first principle that is genuinely
quantum gravitational, rather than just being imported from general
relativity or quantum theory.

But if this is to be the
case the true, fundamental statement of the holographic principle must
be made in the language of some background independent quantum theory 
of cosmology.  This is likely to be phrased in a different language than
its semiclassical formulation, for the same reason that the laws of 
thermodynamics are expressed in very  different language when expressed 
fundamentally in quantum statistical mechanics than they are when one 
first meets them as characterizations of the thermodynamic limit. 
The problem is then to discover what features of the entropy bounds 
and holographic principles so far discussed might be artifacts of the
semiclassical limit, and to separate these from the principle's true
content.

In this paper a line of reasoning is presented, which leads to the 
identification of a form of the holographic principle that can survive passage
to a background independent quantum theory of cosmology.  This is 
called the weak holographic principle\cite{screens}; it is 
both logically weaker and conceptually more radical
than the forms of the principle originally contemplated in 
the literature.  It is logically weaker in that it makes no assertion 
as to a relationship between a bulk and a boundary theory.
As has been found also by
Fischler and Susskind\cite{willylenny} and 
others\cite{others,raphael,raphaelreview}, 
that idea already fails at the 
semiclassical level for cosmological theories. We found in our 
investigations further reasons why such a strong form of the principle
cannot be fundamental.  Instead, the weak holographic
principle comes into a background independent quantum theory of 
cosmology as a framework for that theory's interpretation and 
measurement theory.  Its role is to 
constrain the quantum causal structure of a 
quantum spacetime in a way that connects the geometry of the surfaces 
on which measurements may be made  with a measure of the information 
that those measurements may produce.  In 
this context the entropy bound becomes a definition, by which the 
notion of geometry is reduced to more fundamental notions coming from 
the quantum theory of cosmology. To put it simply, the Bekenstein 
bound is turned on its head and the notion of area is reduced 
fundamentally to a measure of the flow of quantum information.
This form of the principle was first 
suggested in \cite{screens}; the present paper can be taken as an
argument that no form of the principle which is logically stronger, or
conceptually less radical, can survive passage to a background 
independent quantum theory of gravity.

One difficulty of the subject is that different authors have proposed 
different ideas under the name of the holographic hypothesis or
principle.  It is necessary first to bring a bit of order to the 
situation by classifying the different proposals in a way that uses a 
common language and makes 
clear their logical relations to each other.  
To do this we use the language of {\it screens}. In this paper a screen will
always mean an instantaneous, spacelike, two dimensional 
surface\footnote{Or, in $D+1$ dimensions a surface of dimension $D-1$.}
on which quantum mechanical measurements are made.  These will always
be measurements of fields on the surface, which then result in 
information concerning the causal past of the surface.  

To make progress it is first of all 
necessary to distinguish between 
{\it entropy bounds} and holographic {\it principles.}   
The former
are limitations on the degrees of freedom attributable to either the
screens themselves, or spacelike or null surfaces bounded by the 
screens. In the literature entropy bounds are sometimes called holographic 
bounds, but we will stick to the former expression to avoid confusion.
A holographic principle extends an entropy bound by 
postulating a form of dynamics in which the quantum evolution of the spacetime
and matter fields is described in terms of observables measurable on 
the screens. 

We find that the different entropy bounds and holographic
principles that have been proposed fall each into three classes, 
which we call the ``strong''  ``null''  and ``weak'' 
forms\cite{screens}.
The different entropy bounds all postulate that some measure of 
information or of a``number of degrees of freedom''  is
bounded by the area of a screen.  
The strong forms are those that postulate that the bound applies to
the degrees of freedom on a spacelike surface bounded by the screen.
The null forms are those, suggested by Fischler and 
Susskind\cite{willylenny} and put
in a very elegant form by Bousso\cite{raphael,raphaelreview} and 
Flanagan, Marolf and Wald\cite{FMW}, in which there is a 
bound on the number of degrees of freedom of certain null surfaces,
bounded by the screens.  The weak form,  
proposed  with Markopoulou in \cite{screens}, 
postulates only a relationship between the area of the screens and the 
dimension of the Hilbert spaces which provide
representations of algebras of observables on them.

The main conclusion of this paper will
be that the strong entropy bound cannot hold in a 
cosmological theory, and that the null form may only hold in
a semiclassical theory in which quantized matter degrees of freedom 
evolve on a fixed spacetime manifold, but cannot survive the 
quantization of the gravitational field. It appears that 
only the weak form, which as the name suggests is
logically weaker and therefore requires less, may survive in a full
theory of quantum gravity.

The different forms of the entropy bound stem from different 
interpretations that may be given to the Bekenstein bound\cite{bek}.  These,
may be called the strong and weak Bekenstein bounds.  They are presented
in the next section. The distinction is that the weak form bounds only
the information measurable by observers just outside the horizon of a
black hole, while the strong form bounds the total number of degrees 
of freedom measurable in the interior of the horizon.
We find that only the weak form is
required by the usual arguments based on the laws of thermodynamics.
The strong form of the Bekenstein bound follows only if
we add an independent assumption,  which is that the number of degrees 
of freedom measurable on the interior does not exceed those measurable
on the exterior.   We call this the 
{\it strong entropy assumption.}  It must be postulated independently, as 
it does not follow from any argument which involves only measurements 
made exterior to the black hole horizon.
This conclusion has been reached also by Jacobson\cite{ted-bh}.
One of the conclusions of this paper is that the strong entropy 
assumption is false. Among other things it is inconsistent with both inflation and
gravitational collapse.

The strong, null and weak forms of the holographic principle depend
on the corresponding forms of the entropy bounds.  They extend each
of them by giving a framework for dynamics. As only the weak form of 
the entropy bound seems to be possible in a full quantum theory of
gravity, only a weak form of the holographic principle may be true
in such a theory.

The author is aware that this is not a completely welcome conclusion.
In fact, it goes against his own proposal for a bulk to boundary 
isomorphism in quantum general relativity and 
supergravity\cite{linking,hologr,superholo}.  It unfortunately conflicts 
also  with some of the hopes which have been held 
concerning the $AdS/CFT$ correspondence\cite{juan,AdS/CFT}.  It is 
then necessary to discover if there is any conflict between the 
conclusions reached here and the many results which have been found 
which support some version of the $AdS/CFT$ conjecture\cite{AdS/CFT}. 
We find that there is not. This is likely because most of the results so 
far found 
are consequences of much weaker assumptions, which involve 
only the transformation properties of observables under the super-symmetric
extension of $SO(D,2)$.  In fact, Rehren\cite{rehren} has shown 
rigorously that a correspondence will always exist between theories on an 
$AdS_{D}$ background  and conformal field theories on $Mink_{D-1}$, 
subject only to the condition that the latter exist.  The results 
so far found concerning $AdS/CFT$ then may hold as a consequence of this 
theorem.  To the extent that this is true they do not then 
provide any independent evidence for a strong version of the holographic
principle that would go beyond this case.

Does this mean that there is something wrong with the idea that the
holographic principle may play a role in string theory, as was 
suggested by the original arguments for the $AdS/CFT$
correspondence\cite{juan,AdS/CFT}? Certainly, not, what it means is
that if it is to go beyond the level of description in terms of
dynamics of strings and branes in fixed classical background
spacetimes, string theory must be formulated in a background
independent langauge.  Forms of the holographic principle that may
suffice in the context of physics on a single fixed background are likely
to be of limited validity, but the results of our arguments is that
there are forms of the bounds and principle that may hold in a 
background independent theory. In fact, as we argued in \cite{screens},
the weak holographic principle may hold in background 
independent formulations of string theory\cite{mpaper}.

We now give an outline of the paper, emphasizing the logical structure 
of its argument.

This paper is divided into two parts.  The first concerns entropy 
bounds.  In the next section we discuss the weak and strong versions
of the Bekenstein bound and establish the claims made above.

In section 3 we turn our attention to cosmology and find that the weak
and strong Bekenstein bounds each imply a cosmological bound, called 
respectively
the weak and strong cosmological entropy bounds.  As in the 
non-cosmological case, the strong
form cannot be derived without making the strong entropy assumption.

In section 4 we then give five counterexamples to the strong 
cosmological holographic bound.  These are
\begin{enumerate}
	
	\item{}The gravitational collapse problem.
	
	\item{}The inflation problem.
	
	\item{}The wiggly surface problem.
	
	\item{}The two-sided problem.
	
	\item{}The throat problem.

\end{enumerate}

The conclusion is that the strong cosmological entropy bound is 
false.  Since this followed from known physics  plus 
the strong entropy assumption, the likely conclusion is that the latter cannot 
hold in a gravitational theory. 

We then describe, in section 5, the new cosmological entropy bound
proposed by Bousso\cite{raphael}, which we call the null entropy bound. 
It seems to be correct at the
classical and semiclassical level, as a bound on the matter entropy
in a fixed spacetime.  However, to play a role in a quantum theory of
gravity, an entropy bound should extend to a case in which the 
gravitational degrees of freedom are dynamical.  
In section 6 we present two arguments why the null entropy cannot hold 
once the gravitational degrees of freedom are turned on, 
either classically or quantum mechanically.  

The only form of an entropy bound that can then survive at the 
level of a full quantum theory of cosmology is then the weak form.  
This is the conclusion of our discussion of entropy bounds.

The second part of the paper concerns the question of whether, given 
the conclusions reached in the first part, there is any form of a
holographic principle that may hold in a quantum theory of gravity.  
Such a principle must give a 
framework within which to describe the  dynamics of the degrees of 
freedom constrained by the entropy bounds.  Most forms of the 
holographic principle which have been discussed assume the strong form
of the entropy bound.  Dynamics is then formulated in terms of a map 
between the bulk and boundary Hilbert spaces that preserves unitary
evolution.  Since the strong form of the entropy bound seems to 
disagree with things we believe to be true, such a strong form of the
holographic principle is ruled out, at least for the case of 
gravitational theories.

In section 8 we consider this situation carefully, as it is not what
many people's intuition seems to suggest.  We show that there is no 
contradiction with what we know, even taking into account all the
results found concerning the $AdS/CFT$ correspondence.  We also note
that an elegant solution to the black hole information paradox is 
still available.

We then raise the question of whether there might be a weaker form of 
the holographic principle which may still hold.  We consider first, 
in section 9, the question of whether some form of the holographic 
principle may be associated with the null entropy bound.  We reach 
the conclusion that
such a principle may exist, but it must be based on a modification of
quantum theory in which there are many Hilbert spaces, one for each
screen.   

However, if the null entropy bound cannot survive the 
turning on of the gravitational degrees of freedom, neither can the 
null version of the holographic principle. We
are then left with the question of whether there might be a weak 
version of the holographic principle, which would correspond with the
weak entropy bound.  We first, in section 10, discuss the question of
which two surfaces may be screens in such a formulation.  We come to 
the conclusion that none of the possible criteria for distinguishing
screens from other two surfaces can survive passage to the full quantum
theory. Therefor every spacelike two surface may be considered a screen.  This 
opens up the possibility of defining geometry in terms of the 
properties of screens, rather than visa versa.

In section 11 we then list the conclusions of the argument reached till
this point, which then may be considered to motivate and constrain
the possible forms of a weak holographic principle.  One possible 
form of the principle, given in \cite{screens} is then reviewed in 
section 12.  There we also describe briefly 
two independent arguments for a 
weak form of the holographic principle. These come from considerations of 
the role of quasi-local observables in general relativity and, in a 
form made originally by Crane\cite{louis-preholo}, from
relational formulations of quantum cosmology.  

The paper then closes with a short summary of the main conclusions.

\part{ENTROPY BOUNDS}

\section{The weak and strong forms of the Bekenstein bound}

The different possible entropy bounds, as well as the different
possible forms that the holographic principle  might take, have their origin in
the fact that different meanings  may be given to the entropy
of a black hole.  To see this, let us distinguish

\begin{itemize}
	
	\item{\bf The thermodynamic black hole entropy}
	\f
	S_{bh}= {A \over 4 G\hbar}
	\label{bhentropy}
	\ff
	which enters the laws of black hole mechanics.(here $A$ is the
	area of the black hole horizon.)
	
	\item{}{\bf $I^{weak}_{bh}$,  Weak black hole entropy}  This 
	is a measure of how much information an observer
	external to its horizon can gain about its interior, from 
	measurements made outside the horizon.  Besides the mass, angular
	momentum and charges, this includes measurements of the quanta emitted
	by the black hole.  
	
	\item{}{\bf $I^{strong}_{bh}$, Strong black hole entropy} This 
	is a measure of how much information is contained
	in the interior of the black hole.  This can also be expressed
	as the ``number of degrees of
	freedom'' inside the black hole, or the number of distinct ways
	in which it might have been assembled.
	
\end{itemize}

We may note that the generalized second law requires require that
\f
I^{weak}_{bh} \leq S_{bh}
\label{lbhm}
\ff
This is because all of the arguments for it concern exchanges of
matter and radiation between the black hole and observers situated 
outside its horizon.    They do so because they assume that the semiclassical
approximation is valid so that the only way that matter or information
can cross from the interior to the exterior is in the form of 
thermalized Hawking radiation. 

We may note that the number of quanta emitted by a black hole during
Hawking radiation is on the order of (\ref{bhentropy}), this is consistent
with eq. (\ref{lbhm}).  

We are, of course, ignorant of what happens when a black hole
evaporates to any state in which it has a mass of the order of the Planck scale.
The only thing we know with any confidence is that 
the semiclassical approximation breaks down.  To attack this problem
many authors either implicitly or explicitly  make the following
assumption

\begin{itemize}

	\item{}{\bf Strong entropy assumption:}
	\f
	I^{strong}_{bh}=I^{weak}_{bh}
	\label{sea}
	\ff

\end{itemize}	

This is an attractive assumption. For example, it suggests
that there may be a single Hilbert
space, which is a representation of operators at infinity, within which
the full evolution of a system, from prior to black hole collapse
to the aftermath of complete evaporation, may be represented as 
unitary evolution.  However, we should note that the logic is not
symmetric, as there are remnant scenarios under which unitary 
evolution does not imply (\ref{sea}).  Nor can (\ref{sea})
be supported by any arguments for the generalized second law,
as they concern only exchanges of material across the black hole
horizon described in the semiclassical approximation.  Thus, it is
logically possible that (\ref{sea}) is false and that
	\f
	I^{strong}_{bh} > I^{weak}_{bh}
	\label{notsea}
\ff
It is also possible that $I^{strong}_{bh}$ is not even a well defined
quantity.

The arguments which are usually taken as supporting some version
of the holographic hypothesis depend strongly on whether or not
(\ref{sea}) is assumed. To see this, let us run the standard
Bekenstein argument\footnote{This form of the argument is taken from
\cite{pluralism}, but it is due originally to Bekenstein\cite{bek}.  
There is also some confusion because a different bound, 
$S< RE$, where $E$ is the energy of a system was also postulated by
Bekenstein to hold in ordinary quantum field theory, in the absence
of gravity.  The bound we need,(\ref{lbhm}) is logically weaker than 
that, and so arguments against this hypothesis do not necessarily
contradict (\ref{lbhm}).}.

\subsection*{The Bekenstein argument}

Consider a timelike three dimensional region  $\cal R$ of an asymptotically 
flat spacetime $\cal M$, the quantum dynamics of which we wish to study.  
We will assume $\cal R$ has topology 
${\cal R} = \Sigma \times R$, where  $\Sigma$ is a spatial manifold. 
We will restrict attention to the physics within $\cal R$ by
the imposition of boundary conditions 
on $\partial {\cal R} = \partial \Sigma \times R$.  We will 
denote ${\cal S}= \partial \Sigma$.  These will restrict the degrees of
freedom of the gravitational field on the boundary; as a result a 
reduced set of observables will be able to vary at the boundary.

Let ${\cal A}_{\cal S}$ be the complete algebra of the unconstrained observables
on a spatial slice of the boundary, $\cal S$.  This
will have a representation on a Hilbert space ${\cal H}_{\cal S}$.
We will always assume that ${\cal H}_{\cal S}$ is the smallest 
non-trivial representation, i.e. it contains no operators that commute
with the representatives of  ${\cal A}_{\cal S}$.
We will call these the boundary observables algebra and boundary
Hilbert space.  We may assume that among the elements of
${\cal A}_{\cal S}$ are the  Hamiltonian, $H_{\cal S}$ and the areas of regions
$\cal I$ of the boundary $\cal S$, which I will denote $A[{\cal I}]$.
Recall that in general relativity the Hamiltonian
is, up to terms proportional to constraints, 
defined as an integral on the boundary and is thus
an element of ${\cal A}_{\cal S}$.  

Since the system contains gravitation, we may assume that 
among the spectrum of states in ${\cal H}_{\cal S}$ are a subspace
which correspond to the presence of black holes in $\Sigma$.
These are semiclassical statistical states, and
we will assume that their statistical entropies, given by the
dimensions of the corresponding subspaces of ${\cal H}_{\cal S}$
are given by the usual formula (\ref{bhentropy}), 
in the semiclassical limit when their masses and areas are
large in Planck units.

We will consider only systems in thermal equilibrium.  This 
rules out examples from cosmology or astrophysics in which the 
thermalization time or
light-crossing time is longer than the time in which the system will
gravitationally collapse.

The argument is simplest in the case that we assume that
the induced metric on $\cal S$ is spherical, up to small perturbations
corresponding to weak gravitational waves passing through the boundary. The
argument proceeds by contradiction.  We assume that 
the region $\Sigma$ can  contain an object $\cal O$ {\it whose complete
specification in the boundary Hilbert space
$H_\Sigma$}
requires an amount of information 
$I_{\cal O}$ which is larger
than
\f
I_{\cal S}= {A[{\cal S} ] \over 4 l_{Pl}^2}
\ff
which is of course the entropy of a black hole whose horizon just
fits inside of $\cal S$.

Let us assume that initially we know nothing about 
$\cal O$, so that 
$I_{\cal O}$ is a 
measure of the entropy of the system.  
However, with no other information we can conclude
that $\cal O$ is not a black hole,
as the largest entropy that could be  contained in any black
hole in $\Sigma$ is $I_{\cal S}$.  We may then argue, using the
Hoop theorem\cite{hoop} that the energy
contained within $\Sigma$ (as measured either by a quasi-local
energy on the surface or at infinity) 
must be less than that in a black hole
whose horizon has area ${\cal A}[{\cal S}]$.   But this being the case
we can now add energy slowly to the system to bring it up through an
adiabatic transformation to the mass
of that black hole.  By the hoop theorem this will have 
the result of transforming $\cal O$ into
the black hole whose horizon just fits inside the sphere $\cal S$.

This can be done by dropping quanta slowly into the black hole,
in a way that does not raise the entropy of its exterior.  As a
result, once the black hole has formed we 
know the entropy of the system, it is
$I_{\cal S}$.  But we started with a system with
entropy $I_{\cal O}$, which we assumed is larger.  Thus, we have
violated the generalized second law of thermodynamics.  The only way
to avoid this is if $I_{\cal O} <  I_{\cal S}$.  

Since this is a bound on the total information that could
be represented in ${\cal H}_{\cal S}$, we have
\f
\ln Dim \left [ {\cal H}_{\cal S}  \right ] = 
I_{\cal S}= {A[{\cal S} ] \over 4 l_{Pl}^2}
\label{bek}
\ff

We may remark that this argument employs a mixture of 
classical, statistical and semiclassical reasoning.  For example,
it assumes both that the hoop theorem from classical general
relativity applies, in the case of black hole masses
large in Planck units, to real, quantum black holes.  One might
attempt to make a detailed argument that this must be the
case if the quantum theory is to have a good classical limit.
However worthy of a task, this will not be pursued here, 
as it is unlikely that any such argument can be elevated to
establish the necessity, rather than plausibility of the
Bekenstein bound, in the absence
of a complete theory of quantum gravity.  

\subsection*{What does the Bekenstein argument imply?}

We may note that the above argument involves only the weak
form of the black hole entropy, $I^{weak}_{bh}$.  This is because
what is under discussion is the description of the system {\it as given
by states in the boundary Hilbert space ${\cal H}_{\cal S}$}.  This 
is, as emphasized, a representation of the algebra of 
operators ${\cal A}_{\cal S}$ {\it measurable on the boundary.}
This is sufficient to make the argument, as the crucial steps involves
a) use of the hoop theorem and b) adiabatically feeding energy into
the system, both of which concern only measurements or operations
which may be made on the boundary.  The conclusion of the argument
then only concerns the dimension of ${\cal H}_{\cal S}$, and hence
only external measurements.  We may express this by saying that
the argument demonstrates the

\begin{itemize}

	\item{}{\bf Weak Bekenstein bound} Let a system $\Sigma$ be
	defined by the identification of a fixed boundary 
	$\partial \Sigma={\cal S}$,
	and a Hilbert space ${\cal H}_{\cal S}$ be defined as the smallest
	faithful representation
	of the algebra of observables ${\cal A}_{\cal S}$ measurable on the
	boundary only. Either the area 
	$A[{\cal S}]$, is  fixed or is in ${\cal A}_{\cal S}$.
	In the first case,  
	\f
	Dim {\cal H}_{\cal S} \ \leq  \ e^{{A[{\cal S}] \over 4 G\hbar}}
	\label{wbb}
	\ff
	where $G$ is the {\it physical, macroscopic} Newton's constant.
	In the case that $A[{\cal S}] \in{\cal A}_{\cal S}$, 
	the Hilbert space ${\cal H}_{\cal S}$ must be decomposable
	into eigenspaces of $A[{\cal S}]$ such that (\ref{wbb}) is true in each.

\end{itemize}	

Without further assumptions this implies nothing for quantities that 
refer essentially to the ``bulk'' such as ``the number of degrees of
freedom contained in the region $\Sigma$''.  In order that the argument
goes further we may add to it the independent assumption that
(\ref{sea}) holds.  This then does imply the

\begin{itemize}
	
	\item{}{\bf Strong Bekenstein bound.}  Under the same assumptions,
	let ${\cal H}_{bulk}$ be the smallest faithful representation 
	of the algebra of local observables
	measurable in the interior of $\Sigma$.  Then 
	\f
	Dim {\cal H}_{bulk} \ \leq \ e^{{A[{\cal S}] \over 4 G\hbar}}
	\label{sbb}
	\ff
	
\end{itemize}

To summarize, the important points are that the generalized second
law implies only the weak Bekenstein bound, and that the strong
entropy assumption is an independent hypothesis. The logic is then
that
\f
\mbox{Black hole thermodynamics} + \mbox{second law} + \mbox{Hoop theorem}
\rightarrow \mbox{weak Bekenstein bound}.
\ff
and
\f
\mbox{weak Bekenstein bound} + \mbox{strong entropy assumption} 
\rightarrow \mbox{strong Bekenstein bound}
	\label{logic}
\ff

\section{The weak and strong cosmological entropy bounds}

We now turn to the question of whether some form of a holographic bound 
may apply to a cosmological theory in which no boundary conditions
have been enforced.   
Let us consider any closed surface, $\cal S$, which bounds a region
$\cal R$ in a compact spatial slice, $\Sigma$ of a 
cosmological spacetime. No boundary conditions have been imposed on
$\cal S$, thus its interior, $\cal R$ should contain more ``degrees of freedom'' 
than would be the
case were boundary conditions imposed, because boundary conditions 
always act by suppressing degrees of freedom, and hence reducing
the number of classical solutions, in the neighborhood of the
boundary. This means that the above bounds have implications for
the representation spaces of algebras of observables that describe
regions without boundary conditions imposed.

To make this precise, let ${\cal A}_{\cal S}^{free}$ be the total algebra
of observables measurable on $\cal S$, when no boundary conditions
have been imposed, and let ${\cal A}_{\cal S}^{bc}$ be the algebra
of observables
which remain unconstrained when a particular set of boundary 
conditions have been imposed. Let
${\cal H}_{\cal S}^{free}$ and ${\cal H}_{\cal S}^{bc}$ be their
corresponding representation spaces.   Clearly 
${\cal A}_{\cal S}^{bc} \subset {\cal A}_{\cal S}^{free}$, which 
implies that
\f
{\cal H}_{\cal S}^{bc} \subset {\cal H}_{\cal S}^{free}
\ff
This means that 
\f
dim ({\cal H}_{\cal S}^{bc} ) \leq dim ({\cal H}_{\cal S}^{free} )
\ff
We assume that the set of variables which are
fixed by the boundary conditions make up a commuting subalgebra
of ${\cal A}_{\cal S}^{free}$, otherwise they could not all be
imposed at once.  It is also natural to assume that the amount of 
information concerning the state in  ${\cal H}_{\cal S}^{free}$ which
is necessary to fix the boundary conditions is 
proportional to the area $A[{\cal S}]$. It then follows that 
\f
\ln dim ({\cal H}_{\cal S}^{free} )=  \ln dim ({\cal H}_{\cal S}^{bc} )
+ \alpha { A[{\cal S}] \over G\hbar} 
\label{aa}
\ff
where $\alpha$ is some dimensionless constant. We call this the 
{\it  boundary condition area assumption}.

By putting this together
with the weak Bekenstein bound for the system with boundary conditions,
(\ref{wbb}) we find that,
\f
\ln dim ({\cal H}_{\cal S}^{free} )=   
\left ({1 \over 4} + \alpha \right ) { A[{\cal S}] \over G\hbar} 
\ff
Note that this  follows that even though no boundary conditions have been
applied at $\cal S$.  Thus, we have a bound that applies to surfaces 
inside cosmological spacetimes.
\begin{itemize}
	\item{}{\bf Weak cosmological entropy bound.} Let $\cal S$ be
	a spacelike surface of spacetime codimension $2$ that splits
	a complete spacelike hypersurface into two regions, let 
	${\cal A}_{\cal S}^{free}$ be the complete algebra of observables
	measurable on $\cal S$ and let ${\cal H}_{\cal S}^{free}$ be its
	smallest representation space.  Then, 
	\f
	\ln dim ({\cal H}_{\cal S}^{free} )=  C { A[{\cal S}] \over G\hbar} 
	\label{wchb}
	\ff
	for some $\cal S$ independent constant $C$.
\end{itemize}

There is also a strong form of this argument. If we assume the
strong entropy assumption, 
(\ref{sea}), then the same argument leads to 
\begin{itemize}
	\item{}{\bf Strong cosmological entropy bound.} Let $\cal S$ be
	a spacelike surface of spacetime codimension $2$ that splits
	a complete spacelike hypersurface into two regions, let 
	${\cal A}_{\cal S}^{strong}$ be the complete algebra of observables
	measurable on the interior of $\cal S$ and let
	${\cal H}_{\cal S}^{strong}$ be its
	smallest representation space.  Then, 
	\f
	\ln dim ({\cal H}_{\cal S}^{strong} )=  C { A[{\cal S}] \over G\hbar}  .
	\label{schb}
	\ff
\end{itemize}

We again summarize the logic,
\f
\mbox{weak bekenstein bound}+ \mbox{bc. area assumption} \rightarrow
\mbox{weak cosmological entropy bound}
\label{logic2}
\ff

\begin{eqnarray}
\mbox{weak cosmological entropy bound}&+ &\mbox{srong entropy assumption} 
\nonumber \\
&& \rightarrow
\mbox{strong cosmological entropy bound}
\label{logic3}
\end{eqnarray}

\section{Counterexamples to the strong cosmological entropy bound}

Unfortunately, the strong cosmological entropy bound contradicts known
physics.  This is shown by the following five counterexamples.

\subsection*{The  gravitational collapse problem}

Consider a co-moving region $R(\tau )$ in a closed Friedman Robertson 
cosmology,
where $\tau$ is the standard $FRW$ time coordinate.  Let us assume
that at the time of maximum expansion, $\tau_{0}$, $R(\tau_{0})$
contains a uniform gas with entropy $S(\tau_{0}) $, while its boundary has area
$A(\tau_{0})$. If we assume the strong cosmological holographic
bound (\ref{schb}) then $S_{0} < A(\tau_{0}) /4 G\hbar $. However
as the volume of the universes decreases after $\tau_{0}$, so will 
$A(\tau)$. But, by the second law $S(\tau )$ will increase. There
will then be a time $\tau_{1}$ such that $S_{0} = A(\tau_{0}) /4 G\hbar $.
After that the strong cosmological bound will be violated.  
Since the spacetime geometry, and the distribution of gas, are uniform,
the bound cannot be saved by the formation of a black hole.  
A similar problem occurs for boxes of radiation dropped into black 
holes.

Note that this example escapes the conditions of the Bekenstein 
argument because the universe is not asymptotically flat.

\subsection*{The inflation problem}

It is not hard to see that inflation provides counterexamples
to the strong cosmological holographic bound, arising from
the fact that in the aftermath of inflation a universe will
have approximately uniform 
regions exponentially larger than the Hubble scale\footnote{This argument 
has been raised independently in \cite{others}.}.  The  real horizon
size $R_H$ at any given time can then be arbitrarily large compared to the
hubble scale $H$, and still contain entropy created in a single
causally connected region since the initial 
singularity.

To see this 
we follow the exposition of Kolb and Turner, \cite{KolbTurner}. 
We follow a causally connected region which  begins as a
patch of the size of $H^{-1}$ at the time inflation starts, which
is equal to 
\f
H^{-1}= {m_P \over M^2} = R_{initial}
\ff
where $M$ is a mass scale associated with the inflaton potential, which
is between the Planck scale and the weak scale.
The  past lightcone  of this patch will just touch\footnote{so 
that it corresponds to the case Fischler and
Susskind\cite{willylenny} considered.} the initial
surface $t=0$.

There is then a period of inflation, in which the patch expands to
a size $e^N R_{initial}$ which is followed by a period of
reheating, during which it expands by a further factor of
\f
\left ( {M^4 \over T^4 } \right )^{1/3}
\ff
where $T$ is the reheating temperature.  During reheating a
bath of black body radiation is created from the dissipation of
the inflaton field with temperature $T$, after which the inflation
sits in the bottom of its potential and the universe is, to a
very good approximation, spatially
flat.  Agreement with observations seems to require
\f
N > 60
\ff

Just after reheating the region we may try to apply the Bekenstein
bound to the huge bubble that the patch has
grown up to be, which is of radius
\f
R_r = l_P e^N ({ m_p \over T})^{4/3} ({m_p \over M})^{2/3}  
\ff
as space is flat it encloses a volume $8\pi /3 R^3_r$, which
contains an entropy
\f
S_r = {8 \pi \nu \over 3}  T^3 R^3_r = {8 \pi \nu \over 3} e^{3N} 
{ m_P^3 \over M^2 T}
\ff
where
\f
\nu = { \pi^2 \over 40} g^*
\ff
is of order 20 as $g^*$ is of order 100.
If we ask that this entropy be bounded by $1/4$ the horizon area
in Planck units we have
\f
S_r \leq 4\pi^2 R_r^2 
\label{bek1}
\ff
This seems to put a bound on $N$ which rules out
inflation.  This happens because the entropy contained in the
horizon grows
as $e^{3N}$ while its area only grows as $e^{2N}$.  The result is
that (\ref{bek}) implies a strict bound on $R_r$,
\f
R_r \leq { 3\over 2 \nu}({m_P \over T})^3
\ff
which means that the number of efoldings is bounded by
\f
N \leq \ln {3 \over 2\nu} + {5\over 3} \ln {m_P \over T}
- {2 \over 3} \ln {m_P \over M}.
\ff
If we use physically reasonable values for $M$ and $T$ it
is impossible that there were as many as $60$ e-foldings.
Thus, the strong cosmological holographic bound (\ref{schb})
is in conflict with the standard inflationary scenario.

What went wrong?  To see that the problem is inflation we
may note that if we were ignorant of inflation having taken
place, and took the inverse hubble scale $H^{-1}$ for the
horizon size just after reheating instead of the much larger $R_r$ 
the strong cosmological holographic bound (\ref{schb}) yields 
the reasonable statement that,
\f
T \leq \sqrt{2 \over 3 \pi \nu } m_P   .
\ff

So, we may wonder, why isn't the huge region of radius
$R_r$ unstable to gravitational collapse.  It is clearly,
for it has a Schwarzchild radius
\f
R_{Sch} = l_P  4 \pi \nu e^{3N} ({m_P \over M})^2
\ff
Requiring that
$R_r > R_{Sch}$ yields an even stricter bound on the
number of efoldings,
\f
N <  {2\over 3} \ln ( {M \over T}  )-{ 1\over 2} \ln (4\pi \nu ) 
\ff
So the region created by inflation is unstable to gravitational
collapse.  Given any inhomogeneities these
will grow and, if they are large enough, form black holes.
But this is nothing new, it is just the process of galaxy and
structure formation.  Indeed, because $\Omega$ is now very
close to one, large regions of the bubble must be unstable
to the gravitational collapse that must eventually  occur in any
region in which, locally,  $\Omega > 1$.

All of the entropy contained in the
region blown up by inflation corresponds to ordinary thermal
fluctuations in the radiation produced by reheating.  As the
process of reheating is an ordinary physical process, and as
the inflation field may be assumed to have been in a coherent state before
inflation began, we must believe that there are all the degrees of
freedom in that region given naively by the entropy we have computed.
That entropy is, indeed, a measure of how much information
would be needed to determine the precise quantum state which
resulted from the process of reheating.

However, because the causal horizon has blown up to such a
big size from inflation, that information required is much much
greater, for standard inflation models, than the area of the
horizon just after reheating in Planck units.  (recall indeed that
in most standard inflationary scenarios, $N$ is much great than
its minimal value of $60$ and may easily be $> 10^4$.

The inflation problem  shows that there can be a
surface in the universe, the causal horizon, $S_{ch}$, 
whose information content, proportional
to its area, is too small to reconstruct the state of all the
thermal photons in its interior.  Is this a problem for
a consistent cosmological formulation of the holographic
principle?  To answer this we have to ask
what information about the interior may arrive at a surface
at the causal horizon. The key point is that because of the
exponential expansion, an observer there is not able to observe
that thermalization has taken place over all but a small shell
of the interior, in the neighborhood of $S_{ch}$.  For the rest
the observer can see causal effects only from the region prior
to inflation and reheating, when a description in terms of a pure
state is completely adequate.  
This is because, prior to reheating, the state during that era is
very close to the vacuum, and hence can be described with very little
information.

We may contrast this with the information available for 
a surface $S_{oh}$ within the conventional ordinary horizon,
with $r < H$, the Hubble scale.  An observer at such 
a surface sees the region in the interior after reheating and
thus sees a thermal distribution of photons.  But the
region is small enough that enough information is available
on $S_{ohs}$ to reconstruct the states of those thermal photons.

Is there a conflict between these two descriptions?  No, not if
one takes into account two facts.  First, the observer at the
smaller surface must see a mixed state, because the photons
in the interior of $S_{oh}$ will be correlated with photons
in their exterior.  Only from the much larger surface
$S_{ch}$ can an observer reconstruct a pure state, because they
see all the correlations between the thermal photons created
by the inflation and subsequent reheating.  However, it takes
much {\it less} information to describe the pure state than to
describe the thermal state, in the whole of the interior of
$S_{ch}$ because once the quantum correlations
are neglected one must account for all the individual states of
all the individual photons.  

Second, because of causality, the observer at the larger surface
$S_{ch}$ is only able to observe the state in the interior at
a much earlier time, before inflation and reheating, when a pure
state description, requiring much less information, is appropriate.

This examples teaches us that the holographic bound
concerns only the information available
on a surface $S_{ch}$ by virtue of quanta which reach it from
the interior.  This {\it is not the same information}
as would be required to reconstruct the state of the system on
a spacelike surface spanning $S_{oh}$.  They are different because
by causality, the information available on a surface is that information
that can reach the surface by causal propagation of information
from the interior\footnote{One may ask why this example does not 
provide a counterexample to the Bekenstein bound.  The reason is that
a large region of an inflationary universe is excluded because the 
light crossing, and hence thermalization time is long compared to the
time scale for subregions to gravitationally collapse.  These cases
were excluded explicitly in the argument, because the step 
in which one evolves equilibrium states adiabatically by 
slowly dripping in energy cannot be realized.}.  

\subsection*{The wiggly surface problem}
 
Next, we consider  three two-dimensional surfaces in a compact spatial 
slice $\Sigma$.  The first two are 
${\cal S}_1$ and ${\cal S}_2$, with 
${\cal S}_1 \in Int({\cal S}_2)$, where $Int({\cal S}_2)$
is the region in $\Sigma$ to the interior of ${\cal S}_2$.  
For example, these could be
constant $r$ surfaces in a constant $t$ slice of Schwarzchild-De Sitter,
(using standard coordinates)
with $r_1 < r_2$.  In such a case we can choose $A_{1} < A_{2}$, which 
implies that less information could be represented on
$A_{1}$ than on $A_{2}$.    This makes sense because there is
a region between the two surfaces that contains physics
that may be observed by the observer at $S_2$ that is not
observed by the observer at $S_1$.

Now consider a surface $S_1^\prime$, just to the interior of $A_{1}$
which is 
gotten by  displacing $S_1$ slightly into its interior, and then
wiggling it, for example by superposing on it
some set of waves.  The wiggled surface can easily have
area $A_1^\prime > A_2 > A_1$. 
What are we to make of
the apparent fact that the surface $S_1^\prime$, can contain an 
amount of information
greater than the other two?  If ${\cal S}_{1}$ contains all the
information about its interior, then the information coded on
$S_1^\prime$ cannot be greater than that coded on $S_1$. But as
it has a greater area, it seems to have a greater information
capacity.  

The wiggly surface problem 
tells us that the area that is relevant for the measure
of information is not the
actual area of the surface $\cal S$.  Rather it must correspond
to the information reaching $\cal S$ from its interior.  
This can be achieved if
we identify the surface $\cal S$ with a cross-section 
$\sigma ({\cal S})$ 
of a congruence of light rays which intersect $\cal S$.  
We may note that the original arguments of Susskind and others were
phrased in terms of such congruences of light 
rays\cite{lenny-holo,ted-holo}.

This has an important implication.  The bounds on the information on a
screen, $\cal S$,  cannot refer just to that surface.
It must refer instead to the minimal area of cross-sections through
a congruence of light rays that arrive at $S$ from the past.

\subsection*{The two-sided problem}

Consider now a surface ${\cal S}$ of area $A$ and topology $S^{2}$
embedded in a compact
spatial manifold $\Sigma$, which we take to be an
$ S^{3}$.  Then ${\cal S}$ splits the universe into two 
three-balls $B^{\pm}$,such that $\Sigma = B^{+} \cup B^{-}$,
each bounded by a side of $\cal S$,
which we will call ${\cal S}^{\pm}$.  The problem is that
if $ {\cal S}$ is a screen there are actually two
possible holographic descriptions, associated with  
${\cal S}^{\pm}$.  
One  should codes a description of
$B^{+}$, the other should codes a description of $B^{-}$.

Assume that the universe is in a semiclassical state, so that we
may to a reasonable approximation describe the geometry of
$\Sigma$ classically.  Then
consider taking $A$ in Planck units smaller and smaller.  The 
holographic principle must associate to each screen ${\cal S}^{\pm}$ 
a state space ${\cal H}^{\pm}$.
These must have the same dimension,
which is shrinking as $e^{A/4 l_{pl}^{2} }$.  But one of the two  balls,
say $B^{+}$ contains almost the whole universe, while the
other $B^{-}$ contains only a small region.  Since the universe is
classical, it is large in Plank units. We have no problem imagining
that the physics in $B^{-}$ is coded in a state space of
dimension bounded by $e^{A/4l_{pl}^{2}}$, as that is a very small region.
But it seems the physics in $B^{+}$, which is almost the
entire universe must also be describable
in terms of state space of this small dimension.

This seems at first paradoxical.  It seems that this will 
require that an arbitrarily small screen may be required to code
information about an arbitrarily large region.  Is it possible
to resolve this paradox?

It is, if we apply the conclusion of the wiggly surface problem.
We see that what is relevant is not the information that
may reside on a spacelike surfaces spanning ${\cal S}^{+}$
and ${\cal S}^{-}$ but the information reaching those surfaces
transmitted by a congruence of light rays from their pasts.  This
is not necessarily the same thing because to transmit information
the light must be focused on the surface.   We must then consider
the cost in entropy of focusing light from a large universe onto
a small surface.  The apparent loss of information in recording
the holographic image of the large universe on a small surface may
be explained  if the entropy generated (or information required)
by the processes of focusing the light on the surface is large.

Consider a small
surface, $S$, with area of $100 l_{pl}^{2}$ in a large universe
with volume $10^{180} l_{Pl}^{3}$. In order for information about
the state of the whole universe to
arrive at $S$ a congruence of light rays originating all over the
universe must be focused very precisely so that its focal plane
is the surface $S$ at
a fixed time $t$ (measured by a clock at $S$.)   In a universe in
thermal equilibrium, the operation of focusing a congruence of light
rays so precisely, in a manner that compensates for all the
structure in the gravitational field due to the presence and motion of
matter will generate a huge amount of entropy.  
The result\footnote{This can
be made more quantitative, and will be elsewhere.  Note that as
we are discussion the properties of a full quantum theory of gravity,
counter-examples based on classical solutions with isometries are
irrelevant.  It is always possible to find counterexamples to
statistical theorems from examples with non-generic symmetries, 
consider for example the ellipsoid with a point source of light at
one focal point.  It apparently will not stay in equilibrium.} is that 
a great deal of information 
about the universe is then stored, not on $S$, but in
the configuration of matter or lenses that had to be organized
in order to get the light to focus on $S$. 

This will only be unnecessary if the universe is completely symmetric.
However
such a universe will, by virtue of its symmetry, 
contain only a few bits of information\footnote{It might be objected
that there is a limit in which the area, and hence the information
capacity, of the surface is strictly zero.  But this is not true,
under quite generic assumptions in quantum gravity and supergravity
there is a minimal unit of area, which is greater than 
zero\cite{spain,vol1,sn1}.}.

\subsection*{The throat problem}

There are spacetimes, $(M,g)$, in which
the following situation occurs.  $(M,g)$ contains a 
spacelike slices $\Sigma$, with  three 
embedded two dimensional surfaces ${\cal S}_{1},{\cal S}_{2},
{\cal S}_{3}$,
with 
${\cal S}_{3} \in Int({\cal S}_{2} ) \in Int({\cal S}_{1} )$,
but in which their areas satisfy 
$A_{1} > A_{2} < A_{3}$.   This can happen if, for
example, $(M,g)$ is a Kruskal completion of a black hole solution,
${\cal S}_{1}$ is a surface outside the horizon, ${\cal S}_{2}$ is the
throat of the black hole, and ${\cal S}_{3}$ is a two surface which
is at smaller $r$ than the throat, and is topologically
contained in it, but yet has larger area\footnote{This example
is not cosmological, but can easily be made so by inserting the
black hole into a cosmological solution.}.  

There is a large class of such examples in which the Kruskal spacetime
is truncated inside the throat, and a compact region is glued on
containing matter, describing what is sometimes called
a ``baby universe". Such universes are conjectured to arise in a large
class of scenarios in which 
quantum effects lead to an avoidance of the formation of the
singularity.  The problem is that the baby universe is
topologically inside ${\cal S}_2$, but contains $2$-surfaces which 
have a larger area.

To make the problem more worrying we can also imagine that
$A_1 > A_3$.    
In such situations it seems
like  an observer inside the baby universe
at ${\cal S}_1$ can have more information about the contents of the
baby universe than can the observer at ${\cal S}_3$ who is outside the
horizon of the black hole.  This means that the observer at
${\cal S}_3$ may not have enough information to reconstruct the whole
state in the interior of the black hole.  
 
\section{Identifying the wrong assumption}

It is difficult to see how to escape the conclusion that the
strong cosmological entropy bound is false.  One might hope for
escapes from one or two of the counterexamples, but it is
difficult to see how to escape all of them.  The first two are
particularly difficult, as it would be hard to accept a universe
without either gravitational collapse or inflation.  The fact that
the strong cosmological entropy bound prohibits either, in principle,
means that it must be in conflict with the basic principles of
general relativity.

If the strong entropy bound is false, then so must be at least one
of the assumptions that went into its derivation.  This is why we
have been careful to summarize the logic at each step.  The list of
assumptions that went into it are:

\begin{itemize}
	
	\item{}The first and (generalized) second law of thermodynamics
	
	\item{}Classical and semiclassical black hole thermodynamics.
	
	\item{}The hoop theorem.
	
	\item{}The boundary condition area assumption, eq. (\ref{aa}).
	
	\item{}The strong entropy assumption.

\end{itemize}	

Of these, the boundary condition area assumption is a technical 
assumption, that helps make the argument cleanly, but if we had to 
drop it we 
could still construct the counterexamples. They would just take place 
within a box, rather than in a cosmological spacetime and so they 
would then 
contradict the strong form of the Bekenstein bound.  But they would
bite no less in that context.  

There is a great 
deal of evidence for classical and semiclassical black hole 
thermodynamics and we have no independent evidence that the basic 
principles of thermodynamics are not to be trusted in this regime.  
The hoop theorem is also well understood and established.  The only
assumption on this list without independent support is the
strong entropy assumption.  It must then be wrong.  

In fact, as we have emphasized, none of the arguments in the subject
provide any independent support for the strong entropy assumption.
There is then no argument for the validity of anything stronger
than the weak Bekenstein bound and weak cosmological entropy bound.

\section{The  null entropy bound}

We now turn to a different kind of entropy bound
bound, proposed by Bousso\cite{raphael}, following a suggestion of
Fischler and Susskind\cite{willylenny}.  They proposed
that the bound  
restricts the information, or number of degrees of freedom
on null, rather than spacelike surfaces bounding the
screens.  

This principle may be stated as follows.

\begin{itemize}
	
	\item{}{\bf Null entropy  bound (Bousso)}.
	We fix a spacetime manifold $({\cal M},g)$ on which a quantum field
	theory has been defined.  A {\it  screen} $\cal S$ will be 
	an oriented two 
	dimensional spacelike surface, possibly open.  We then consider one 
	of the four congruences of null geodesics which leave the screen
	orthogonally, either to the future or the past, and to the left  or
	right of the screen.  These may be labeled $L^{\pm}_{l,r}$.
	We call each a {\it light surface} associated 
	to $\cal S$.
	
	\item{}Each of the four  light surfaces $L^{\pm}_{l,r}$ may contain
	a subsurface, $\cal L$ which satisfies the following condition:
	The expansion of null rays
	$\theta$ (in the direction going away from $\cal S$)
	is non-positive at each point of $\cal L$. If the boundary
	of $\cal L$ contains $\cal S$ then we call $\cal L$ a {\it light
	sheet} of $\cal S$. The boundary of $\cal L$ will generally contain,
	besides $\cal S$, a set on which the condition $\theta \leq 0$ fails to 
	hold, either because of the existence of crossing points or caustics,
	or because the lightsheet intersects a singularity of the spacetime.
	
	\item{}Now, let $\tilde{s}^{a}$ be the entropy current density of matter, so that 
	\f
	S[{\cal L}]=\int_{\cal L}d^{3}x_{a} \tilde{s}^{a}
	\ff
	is the entropy crossing the light sheet.  
	Then the null entropy bound is 
	\f
	 S[{\cal L}] \leq { A[{\cal S}] \over 4 G\hbar }
	\label{mschb}
	\ff
	
\end{itemize}

Before turning to its implications, we should 
mention a possible counterexample, which was proposed by 
Lowe\cite{lowe}, as its refutation shows the subtlety of the null 
entropy bound.  Consider a box containing a Schwarzchild black hole and thermal
radiation, which are in equilibrium at a temperature $T$. 
If the box is small enough the ensemble including the black hole
has positive specific heat, so the equilibrium is stable.
Let us then
consider a spherical spacelike two surface, $\cal S$ which is a slice of the
horizon $H$ of the black hole.  Lowe suggests that the horizon 
$H^{+}$ to the future of $\cal S$ should be considered a light sheet of $\cal S$. 
But as the geometry is static, $H^{+}$ has no boundary besides $\cal 
S$, so that $\int_{\cal H^{+}}d^{3}x_{a} \tilde{s}^{a}$ will diverge, 
since $ \tilde{s}^{a}$ is also constant in equilibrium.
Thus, (\ref{mschb}) is apparently violated.

The problem, as pointed out by Bousso\cite{raphael,raphaelreview} is that in 
determining the actual light sheets of $\cal S$ we cannot use the 
static geometry, as that is just an averaged description of the actual
spacetime geometry. As in any case in which the second law is 
evoked in a statistical system, we must be careful to 
take the thermal fluctuations around equilibrium into account. They 
cause small fluctuations in the spacetime geometry, the result of 
which is that the actual light sheets $\cal L$ of $\cal S$ will not
coincide with $H^{+}$, when the latter is defined in terms of the 
average, static geometry.  Instead, the small fluctuations will cause 
parts of the real light sheet to deviate either inside or outside of the 
averaged horizon.  Those that fall inside will shortly hit the 
singularity (or else, if the singularity is avoided by a bounce, cross, 
causing caustics.) 
Those that deviate away from the horizon no longer satisfy the 
condition that $\theta \leq 0$. Thus the real light sheets will have
outer boundaries.   Bousso then argues in \cite{raphaelreview} that the
bound will be satisfied.

We may note also that the possible counterexample could be avoided if 
one only required that the light surface satisfy $\theta <0$, so as 
to rule out the marginal case $\theta=0$.  Bousso chooses not to do 
this as that would eliminate the important example of static black 
hole horizons.

Finally, we may note that a 
proof of a closely related conjecture has been given in \cite{FMW}.

As a result, it appears, at least as of this writing, that Bousso's
null entropy bound agrees with everything we know.  It then may be
considered to be a useful, and surprising feature of general relativity
at the classical and semiclassical level.  

\section{Could the null entropy bound extend to quantum gravity?}

While the preceding is very satisfactory, we must note that 
the null entropy bound is formulated in terms of the behavior of
the entropy of matter, on a fixed spacetime background. This is 
already interesting, but for the possible application to quantum 
gravity we should ask more.  This is because 
general relativity is a dynamical theory, with its own degrees of 
freedom.  We would then like to know if there is an extension of the
bound, even at the classical level, which applies not just to a single
spacetime, but to a family of spacetimes which differ by the 
amplitudes of gravitational waves which may be present in the region 
containing a screen or a light sheet.  If this were the case then a 
corresponding result would be more likely to hold in quantum gravity, 
in which there is no fixed spacetime.  

Another reason to demand this is that supersymmetry, which seems to be
required for the perturbative consistency of quantum gravity, tells us 
that the distinction between the matter and gravitational degrees of 
freedom is gauge dependent, and hence not physically meaningful.  
Furthermore, in perturbative string theory, which seems necessary for 
perturbative quantum gravity, both the matter and gravitational degrees of 
freedom arise from excitations of more fundamental degrees of freedom.
A bound which requires a strict separation of matter and gravitational 
degrees of freedom cannot then be formulated in a manner consistent 
with local supersymmetry and is hence unlikely to extend to 
supergravity or string theory.

We then investigate, in this section the question of whether there
might hold an extension of the null entropy bound that would hold in
either of the following cases: i) as a statement on the phase space of
general relativity, which allows the fluctuations in the gravitational
degrees of freedom to be turned on, ii) in a quantum theory of gravity 
or iii) in a locally supersymmetric theory.  While we do not decide the
question, we find that there are two worrying issues, which we now 
describe.

\subsection*{First problem: Including the gravitational degree of freedom}

When the gravitational degrees of freedom are turned on we
face a paradox because the light sheets, $\cal L$ on which
the entropy is measured, depend for their definition on the actual values of the 
gravitational degrees of
freedom.  This dependence is not weak or gradual, as the positions
of the singularities of $\sigma$, and hence the location of the 
boundaries that define ${\cal L}$ depend non-linearly on
the values of the gravitational degrees of freedom.  
Thus, even
in classical general relativity, it is difficult to know what would
be meant by the null entropy bound once the gravitational
degrees of freedom are turned on.

The point may be put the following way. Consider a fixed spacetime
$({\cal M},g_{ab})$, which has a Cauchy surface, $\Sigma$, 
which has embedded in it a screen $\cal S$ which has  a future light 
sheet $\cal L$. $\cal L$ is then to the future of $\Sigma$.  
Now, consider a one parameter family of metrics 
$g_{ab}^{s}$, such that $g_{ab}^{0}=g_{ab}$
which, for $s \neq 0$  differ from $g_{ab}$ in a region  
$\cal F$ which is to the causal future of a region $\cal R$ of 
$\Sigma$, not containing $\cal S$.  For each $s$ one can identify the
light surface formed by the future null congruence $L(s)$ from $\cal 
S$ such
that $L(0)$ contains $\cal L$.  In fact, for each $s$ there
will be a light sheet ${\cal L}(s) \subset L(s)$.  Using this one can identify 
for each $s$ a region $U(s) = L(s) \cap {\cal F}$. Let us pick
$\cal R$ such that at $s=0$, $U(0) \subset {\cal L}$.  

Now, the notion of a light sheet would be preserved under variations
in the gravitational degrees of freedom were it the case that for 
all $s$, and all such one parameter families,
$U(s) \subset {\cal L}(s)$. However it is easy to see that this is 
not the case\cite{rezageorge}.  The reason is that one can always find 
one parameter 
families $g_{ab}^{s}$, specified by initial data in $\cal R$ such that, 
for a finite  $s$, all of $U(s)$ will not be in the future lightsheet
$L(s)$.  The reason is that the 
gravitational radiation will  induce caustics to form in $U(s)$
causing $\theta$ to be positive on some part of $U(s)$.  

This means that there is no definition of a light sheet which is 
independent of the initial data in a region $\cal R$ of a Cauchy 
surface containing a screen $\cal S$, even if $\cal R$ does not 
contain $\cal S$.  The null entropy bound may hold for each 
light surface ${\cal L}(s)$ in each spacetime $g_{ab}^{s}$ but there
is no extension of the result which holds on the space of solutions or
initial data of a spacetime, even when the degrees of freedom are 
restricted to vary in regions that do not include the screen.  

What this means is that one cannot extend Bousso's bound to include
the gravitational degrees of freedom in any way which involves
defining the entropy of the gravitational degrees of freedom in terms
of  statistical ensemble of states or histories.

But if this is the case then it is hard to see how there could be
an extension of the null entropy bound either to quantum gravity, in 
which case there is no fixed classical spacetime, or in supergravity, 
in which there cannot be an invariant distinction between gravitational
and matter degrees of freedom.

\subsection*{Second problem: measurability}

Another aspect of the problem just discussed is that once the
gravitational degrees of freedom are turned on, either classically
or quantum mechanically, whether a particular null surface $\ L$ is
a lightsheet of some screen or not depends on the values
of the degrees of freedom.  This means that there are three choices:
I) find a formulation of a cosmological entropy bound that does not
require the identification of a lightsheet on which 
$\theta \leq 0$, II)  try to formulate the condition as an operator 
equation or III) try to formulate it in terms of expectation values.

The first possibility leads us to the weak entropy bounds, in 
which the lightsheet plays no role.  In case II
one must find
a set of commuting operators in quantum gravity which are sufficient 
to
define the notion of a light sheet and apply it to their
eigenvalues.  
We must then ask whether the  uncertainty principles which arise from the
commutation relations of a quantum theory of gravity allow
the simultaneous measurement of quantities that must be known to apply the bound.
To investigate this  we shall 
assume the standard equal time commutation 
relations\cite{abhay,action,sam}
\f
[ A_{a}^{i}(x,t) , \tilde{E}^{b}_{j}(y,t)  ]= \delta^{3}(x,y) \delta^{b}_{a}
\delta^{i}_{j}
\ff
where $A_{a}^{i}$ is the self-dual connection and $\tilde{E}^{b}_{j}$
is the dual of the  pull back of the self-dual two form, $\Sigma$ in any
spacelike surface (and all other commutators vanish).  We may note that these hold in a large class
of theories, including all the extended supergravities, as they
arise from the generic form\cite{action}
\f
I= \int_{\cal M}\Sigma^{AB} \wedge \dot{A}_{AB} + \ldots
\ff
It is difficult to imagine that these do not hold in the effective
field theory which is the low energy limit of string theory or 
whatever the true quantum theory of gravity is.  

It is not hard to show that,

\begin{itemize}
	
	\item{}If $\theta^{\pm} (s)$ is the expansion of the future and past
	going null geodesics normal to a two surface $\cal S$ at a point
	$s\in {\cal S}$ and $A[{\cal S}]$ is the area of $\cal S$ 
	then\cite{jerzy-personal}
	\f
	\left [ \theta^{\pm} (s), A[{\cal S}] \right ] =0
	\ff
	
	\item{}Let $(s,u)$ be coordinates on a null surfaces, $L$, generated by the 
	null geodesics leaving $\cal S$ orthogonally, 
	where $s$ labels a congruence of null geodesics
	and $u$ is an affine parameter along each geodesic such that
	$\cal S$ is defined by $u=0$.  If
	$\theta^{\pm} (s,u)$ is the expansion of the congruence
	at $(s,u)$
	then, for $u \neq 0$ we have,
	\f
	\left [ \theta^{\pm} (s,u), A[{\cal S}] \right ] \neq 0
	\ff
	\f
	\left [ \theta^{\pm} (s,u), \theta^{\pm} (s)   \right ] \neq 0
	\ff
\end{itemize}

Thus, there does exist a basis which comes from simultaneously diagonalizing
the expansion of the congruence of null rays at a screen, and 
its area. But in such a basis the operators 
$\theta (s,u)$ off the screen are not diagonal.   
Thus, we cannot identify the outer
boundary of the lightsheet in any basis in which we can identify
a screen and measure its area.
This suggests that there cannot exist an operator form of the 
null entropy bound.  

The remaining possibility is to formulate the condition for a screen 
in terms of expectation values.  Presumably this should be possible in 
the semiclassical limit, otherwise the null entropy bound could not
be true in that limit. I am not aware of any proposal to implement 
it beyond the limit.  One way to see why this is unlikely to work is
to discuss how it would have to work in a path integral formulation 
of  the theory.  

Let us first note that to be relevant to the null entropy bound, a 
histories formulation will have to be formulated in terms of causal
histories, as there are no analogues of light sheets in Euclidean 
metrics.  Fortunately, there are now non-trivial 
proposals\cite{FM2,fmls1,tubes,pqtubes,sameer,withstu,CW} and 
results\cite{AL,ANRL} concerning formulations of quantum gravity in 
terms of causal,lorentzian path integrals.  In such a causal histories
formulation, each history in the sum over histories
comes with its own causal structure.  The problems we are discussing can
then be stated as follows: it may be possible to pick out consistently
a set of histories in which the area of a preferred family of
surfaces and the expansions of null geodesics at those surfaces
are given and fixed.  But in these histories the properties of the
light surfaces generated by following null geodesics from the
screens cannot be controlled, and will fluctuate as the sum over
histories is taken.  Since caustics and other singularities are
generic for such surfaces, the observer at the screen will be
unable to  control or measure any variables at the screen to prevent
the formation of singularities in the light surfaces of the histories.

Another way to say this is that the degrees of freedom of the light
surface include components of the metric and connection on the
light surface itself.  Singularities and caustics will form
for generic values of these parameters, but where they form 
varies as the degrees of freedom fluctuate.  One cannot then
consistently count the number of degrees of freedom on the
non-singular part of the light surface, this involves a logical
contradiction since the presence and location of the singularities
itself depends on the values of the fluctuating degrees of freedom.

Before closing this discussion, we note that our argument does not 
exclude one radical possibility, which is that there is an entropy
associated with single classical configurations of the gravitational
field.  This has been suggested by Penrose\cite{roger-weyl}, and there
are some old arguments for it, which rely on the impossibility of 
either building a container to constrain gravitational radiation, inducing
a gravitational ultraviolet catastrophe or measuring the pure
state of gravitational radiation\cite{graventropy}.   In the present
context this would suggests that 
$\int_{\cal L} \sigma^{ab}\sigma_{ab} $, where $\sigma$ is the shear of the null
congruence, might be taken as a measure of 
the gravitational entropy on a lightsheet $\cal L$ in a single classical 
spacetime.  

This is an intriguing possibility, which deserves investigation.  It 
does not affect the following considerations, as the difficulties we 
find with a null form of the holographic principle would not be 
lessened.

\part{HOLOGRAPHIC PRINCIPLES}

A holographic principle is meant to be a formulation of the dynamics of
a quantum theory in terms of its screen, or boundary hilbert spaces.
A holographic principle requires some form of an entropy bound, but it
requires also that the dynamics of the theory can be 
formulated entirely in terms of the degrees of freedom measurable on 
the screen.  

There are different kinds of holographic principles, corresponding to
the different possible kinds of entropy bounds.  We consider in turn, 
strong, null and weak forms of the holographic principle.

\section{The strong holographic principle}

The classic formulation of the strong holographic principle is meant 
to apply to the case we discussed in section 2.
We have a quantum or classical spacetime, $\cal M$, with a 
boundary $\partial {\cal M}= R\times {\cal S}$, where $R$ corresponds
to the time coordinate. One then postulates boundary and bulk
algebras of observables, ${\cal A}_{S}$ and ${\cal A}_{bulk}$
and  ${\cal H}_{S}$ and ${\cal H}_{bulk}$ as in section 2.  In 
addition one specifies on each hilbert space a hermitian hamiltonian
$h_{\cal S}$ and $h_{bulk}$.  In a gravitational theory this may
require the specification of gauge conditions on the boundary, in this
case there is a family of bulk and boundary hamiltonians which depend 
on the gauge conditions.  The principle is then formulated as follows

\begin{itemize}
	
	\item{}{\bf Strong holographic principle}  There is an isomorphism
	\f
	{\cal I}: {\cal H}_{bulk} \leftrightarrow {\cal H}_{S}
	\label{shp}
	\ff
	such that ${\cal I}\circ h_{bulk}=h_{\cal S}$.  
	
\end{itemize}

Can this principle be satisfied?  There are two cases: gravitational 
theories and non-gravitational theories. 

\subsection*{Failure of the strong holographic principle in 
gravitational theories}

By a gravitational theory we mean here theories in which the 
gravitational degrees of freedom are allowed to fluctuate, so that
any principle must hold for all the possible initial data
that the theory allows.  This will be specified 
classically or semiclassically by initial data on $\Sigma$ 
or quantum mechanically by the specification of a state in ${\cal H}_{bulk}$.
We do {\it not} mean quantum field theory on a particular fixed spacetime 
metric $g_{ab}$ on $\cal M$.

There is some evidence that the strong holographic principle may hold 
in a quantum theory of gravity.  One piece of evidence comes from 
canonical quantum general relativity, with a non-zero cosmological
constant, and certain boundary conditions, called the Chern-Simons
boundary conditions\cite{linking,hologr,superholo}.  In this case the 
boundary Hilbert space is found to be of the form
\f
{\cal H}_{\cal S}= \sum_{a} {\cal H}_{\cal S}^{a}
\ff
where $a$ is an eigenvalue of the area operator, which is known from
\cite{spain,vol1} to have a discrete spectrum. Each of the eigenspaces
${\cal H}_{\cal S}$ is a space of $SU_{q}(2)$ intertwiners on a 
punctured $S^{2}$ where the labels on the punctures, which are
taken from the representations of $SU_{q}(2)$ are related to 
the area\cite{linking,hologr}.  The level $k$ is related to the
cosmological constant by\cite{linking}, $k=6\pi /G^{2}\Lambda$.  
The dimension of the space of intertwiners does satisfy  eq. ( \ref{bek}),
with a renormalization of Newton's constant defined by $G_{ren} = 
cG_{bare}$, with $c=\sqrt{3}/\ln (2)$. 
Thus, it is clear that the weak Bekenstein bound is satisfied.

There are also some results concerning the bulk Hilbert space.
Before the hamiltonian constraint is imposed, an infinite dimensional
space of bulk states can be identified for each $a$, which has an
orthonormal basis given by the distinct embeddings (up to 
diffeomorphisms) of 
quantum spin networks in the bulk,  whose edges meet the boundary at
the punctures.  One can then show that there is, for each set of
punctures, a finite dimensional space of solutions to the Hamiltonian
constraint which is isomorphic to the corresponding boundary Hilbert 
space\cite{linking}.  These are constructing by moding out the infinite 
dimensional
kinematical Hilbert spaces by a set of equivalence relations which
generate the recoupling identities of quantum spin networks. It is 
known that for a certain class of states these recoupling identities
realize the action of the Hamiltonian constraint\cite{linking}.  

In this case the strong entropy assumption then comes down to the
conjecture that these provide a complete set of solutions to the
Hamiltonian constraint in the bulk.  There is presently no evidence
either way on the correctness of this conjecture.  It is attractive to
argue that if it is true we have quantum general relativity in this
particular case expressed in closed form as a theory which satisfies
the strong holographic principle.  We may also note that these results 
all hold for both the euclidean\cite{linking} and 
lorentzian\cite{hologr} cases as well as for 
supergravity\cite{superholo}.

However, if we accept the conclusion of the previous arguments, then 
this conjecture must in fact be false.  We have found instead that 
for  the case of a gravitational theory  the strong 
holographic principle cannot hold unless the boundary of the
spacetime has either infinite or indeterminate area.  The reason is that, as we 
have shown, the strong entropy assumption, and the strong forms of the
Bekenstein bound and the cosmological entropy bounds all fail.  As a
result we cannot assume that the bulk and boundary Hilbert spaces have
the same dimension.  However, we have also shown that the weak 
Bekenstein and weak cosmological entropy bounds hold, which means 
that ${\cal H}_{\cal S}$ is finite dimensional or, generically, is 
composed of finite dimensional subspaces, which are the diagonal 
sectors of the area $A[{\cal S}]$. 

Since no bound has been found to hold which restricts the dimension of
${\cal H}_{bulk}$ there are two possibilities, either it is infinite
dimensional, or it does not exist.  In the latter case there is nothing
for an isomorphism to map the boundary state space to.  If it is 
infinite dimensional it could only be mapped to the boundary which has 
either indeterminate or infinite area.  Thus we conclude that 
a form of the strong holographic principle could only hold in those 
cases.  

Let us now consider the case of indeterminate area more closely.  Let 
us consider the Schroedinger picture operators, defined in terms of 
the time at the boundary. It is clear that by causality $\hat{A}[{\cal S}]$, the 
operator that measures the area of the boundary must commute with 
$h_{bulk}$, as the latter is a function only of degrees of freedom in 
the bulk of  
$\Sigma$ which are causally unrelated to degrees on $\cal S$. Another 
way to say this is that where we  must be able to move the boundary 
locally, thus changing its area, without affecting the physics in 
regions of the bulk causally disconnected from the events of moving 
the boundary.  Since $[\hat{A}[{\cal S}], h_{bulk}]=0$, we must be
able to construct projection operators in the bulk, $\hat{P}_{a}$ 
corresponding to every eigenvalue, $a$ in the spectrum  of $\hat{A}[{\cal S}]$,
and define the restricted Hamiltonian $h^{a}_{bulk}= \hat{P}_{a}h_{bulk}\hat{P}_{a}$.
By causality it must then be the case that the strong holographic principle work between 
the corresponding subspaces ${\cal H}^{a}_{\cal S}$ and ${\cal H}^{a}_{bulk}$
for {\it each} value of $a$.  

But now we can apply the argument for the case of a finite area 
boundary.  Since there is no bound on the dimension of ${\cal H}^{a}_{bulk}$,
it must have infinite dimension, thus it cannot be isomorphic 
to ${\cal H}^{a}_{\cal S}$.

This leaves only the case that the boundary has infinite 
area.  But in this case there cannot be a cosmological version 
of the principle, as generic spatial regions in generic cosmological 
solutions have finite area boundaries.  Thus, at best, the strong 
holographic principle could only apply to the case of non-compact 
spacetimes with boundary.  

This may be satisfactory, but it comes with a price, which is that we 
will not be able to apply the principle to any case in which the 
boundary is moved inside the non-compact spacetime to coincide with a 
finite area surface.  
This means that for any such surface, labeled again by its area, $a$ 
the holographic correspondence (\ref{shp}) 
will  map all but a finite dimensional subspace of ${\cal H}^{a}_{bulk}$ 
to degrees of freedom that are contained within  ${\cal H}^{\infty}_{\cal 
S}$, but are not representable within ${\cal H}^{a}_{\cal S}$.  This 
must hold for any finite $a$.  It means that for no finite $a$ can there 
be any correspondence between ${\cal H}^{a}_{bulk}$ and 
${\cal H}^{a}_{\cal S}$, as almost all of the information in the 
former is not representable in the latter.  This is counterintuitive, 
it means no matter how far we move the boundary out, the 
representation space of the boundary observables do not capture most 
of the information about the bulk observables, so long as the area of 
the boundary is finite.  

This would be very disappointing, what it really means is that there 
is no way going to an infinite boundary can save the situation, once 
it is realized that for any finite area boundary there can be no 
holographic isomorphism (\ref{shp}).  Thus, we conclude that there 
cannot be an implementation of the strong holographic principle in a 
gravitational theory.

\subsection*{Realization of the strong holographic principle in 
non-gravitational theories}

It is surprising, and striking, that in spite of its failure for
gravitational theories, there are realizations of the strong 
holographic principle for non-gravitational theories. These occur in 
a special case which is Anti-DeSitter backgrounds in 
$D+1$ dimensions\cite{juan,AdS/CFT,rehren}.  
In these spacetimes the asymptotic boundary is 
timelike and is in fact conformally compactified Minkowski spacetime 
($CM$)
in $D$ dimensions. The existence of such a correspondence was 
conjectured first by Maldacena in a string related 
argument\cite{juan,AdS/CFT}, but has 
since been shown to hold quite generally for non-gravitational 
theories on AdS backgrounds\cite{rehren}.  There is in fact a rigorous 
theorem in axiomatic
quantum field theory that shows that gives such an isomorphism
for generic field theories on $AdS$ spacetimes\cite{rehren}.  

The reason behind this correspondence is clearly that $SO(D,2)$ acts as
the symmetry group on $AdS_{D+1}$ and as the conformal symmetry group
of $CM_{D}$.  As a result one can establish an isomorphism (\ref{shp})
for general quantum field theories on $AdS$ backgrounds.  

We can also see why the argument given just above is superseded in the
$AdS$ case. A key fact is  that $AdS$ spacetime has no Cauchy surface.  The
reason is that the evolution in the bulk requires the specification
of data on the timelike asymptotic boundary of the spacetime. If
the boundary fields are not specified there is
no deterministic evolution for the bulk degrees of freedom. As
a result the boundary degrees of freedom are part of a complete
specification of the dynamics of the bulk theory.  This makes it
less surprising that the dynamics can be reduced to a description
of boundary degrees of freedom, in this case the bulk to boundary map
is plausibly a reduction to the data necessary to determine a solution,
and may play a role similar to that of the map that relates a solution 
to boundary data in spacetimes with Cauchy surfaces.  This is very different 
from what happens in asymptotically flat spacetimes in which only the only
quantities measurable at spatial infinity are a finite set of 
conserved quantities.  

The key question is then whether there are conformal quantum field 
theories for general $D$ on ${CM}_{D}$.  There are certainly free field
theories, for which the correspondence holds\cite{rehren}.  It may then
be expected to hold also for interacting theories in those special
cases in which there is a conformal quantum field theory on ${CM}_{D}$.
There is evidence that one such case is $N=4$ supersymmetric 
Yang-Mills theory for $D=4$\cite{juan,AdS/CFT}.  By the general arguments of 
\cite{rehren} one would expect there to exist on $AdS$ a supersymmetric 
theory, whose spectrum transformed under a supersymmetric extension 
of $SO(4,2)$, with 16 supercharges.  There is a great deal of 
evidence that this is the case, at least in the limit $N\rightarrow 
\infty$. In this case the theory appears to be the weak coupling limit of 
supergravity compactified on an $AdS_{5}\times S^{5}$ background.

\subsection*{The big question}

We then have the following question.  We have argued that the strong 
holographic principle cannot hold in a gravitational theory. It can 
hold in a quantum field theory on a fixed background, and indeed in 
the particular case of $AdS$ spacetimes it seems to be a generic 
feature.  But it should be expected to break down as soon as the 
gravitational degrees of freedom are turned on\footnote{The same 
questions can also be asked in the $2+1$ dimensional case, where the
$AdS/CFT$ correspondence has also been worked out\cite{AdS/CFT}. However, 
given that quantum gravity and supergravity in $2+1$ dimensions 
are topological quantum field theories, there may be little to learn
from this case that is generally useful.  TQFT's are by 
definition theories whose observables and states are defined on 
boundaries of the spacetime.}.  More precisely, we 
expect our first two counterexamples to arise as soon as either gravitational 
collapse or inflation could occur in regions of the bulk. 

At the same time, in the particular case of $AdS_{5}\times S^{5}$ the 
isomorphism seems to exist and the bulk theory is then the weak field 
limit of a gravitational theory.  What then happens when the 
gravitational constant is turned up, so that the gravitational degrees 
of freedom are excited? This is the big question.  There seem
to be three possibilities:

\begin{itemize}
	
	\item{}{\bf 1} Something is wrong with the above reasoning, at least 
	in the case of supersymmetric theories.  
	
	\item{}{\bf 2} In supergravity or string theory on spacetimes which are
	asymptotically $AdS_{5}\times S^{5}$ the counterexamples cannot 
	arise.  This means that there cannot be small black holes that form 
	from gravitational collapse and there cannot be any possibility of 
	choosing the initial conditions in the interior so as to drive the 
	theory into an inflating phase.
	
	\item{}{\bf 3} The correspondence holds at the level of the background 
	dependent quantum field theory defined on $AdS_{5}\times S^{5}$
	by the weak coupling limit of supergravity or string theory, but
	breaks down as soon as the gravitational constant or the initial
	data is large enough that strong gravitational fields can arise.
	
\end{itemize}	

The first possibility is of course always there, this is why we have been very 
careful to keep track of the logic leading to the conclusion that the
strong entropy assumption, strong Bekenstein bound and strong
cosmological entropy bounds must all be false in gravitational 
theories.  If there is an error it must be either in the reasoning or 
in the unexpected failure of one of the other assumptions listed in
section 5.

The second possibility seems unlikely, and in any case were it true it 
would mean that this particular case is non-generic in ways that 
suggest it is not a very good example of a quantum theory of gravity.

We must then ask if there are any results that contradict the third
possibility.  At of this writing all of the results found which 
support the conjecture in the $AdS_{5}\times S^{5}$ case relate 
boundary observables of supergravity on that background to expectation
values of the supersymmetric Yang-Mills theory.   While the 
construction of representatives in the Yang-Mills theory of bulk 
observables in the bulk theory have been discussed, there is so far 
no calculation which gives a non-trivial test of these correspondences.
It is also the case that most, if not all, of the calculations of 
$N$-point functions which support the conjecture are in any case 
forced by the action of the super-symmetry group.  It then seems to 
be the case that even if it disagrees with some interpretation of the 
conjectured correspondence, there are no actual results which so far
contradict possibility {\bf 3.}

There is a final remark which is consistent with this third 
possibility which is the following.  Gravitationally bound systems 
including black holes have generically negative specific heat.  
However, the positivity of the specific heat for an equilibrium 
ensemble is guaranteed for any system defined by a partition function.
In particular, the thermal quantum field theory gotten by raising the 
temperature in the 
$N=4$ supersymmetric Yang-Mills theory is defined by a partition
function.  Therefor all equilibrium configurations will have
positive specific heat.  

We can then ask how configurations such as a system of planets or 
small black holes in the bulk of the AdS spacetime are to be represented 
in terms of states of the $N=4$ supersymmetric Yang-Mills theory. 
There are two possibilities: this can be done, but involves 
configurations that are sufficiently far from equilibrium in the 
Yang-Mills theory that they cannot be described by a partition 
function.  Or, the correspondence breaks down as soon as 
gravitationally bound states of the bulk theory arise whose 
statistical ensembles have negative specific heat.  

As a final remark, we note that $S$ duality is still a conjecture,
outside of the $BPS$ sector of either $N=4$ superYang-mills theory
or string theory.  Thus, if the isomorphism (\ref{shp}) fails beyond the 
$BPS$ sector there is nothing that constrains $S$-duality to hold
on both sides of it.

\subsection*{What about the black hole information paradox?}

One reason that the strong holographic principle has been advocated
by some people is that it guarantees a solution to the black hole
information paradox.  Thus, one can wonder if there is an 
independent argument for the strong holographic principle which 
follows from the possibility that it is necessary to give a consistent
resolution of the black hole information paradox.

The answer is negative, because a large part of the black hole 
information paradox depends on the strong entropy assumption, which 
we have found is false.  Once it is realized that the strong entropy
assumption is false, there is no reason to presume that the amount of 
information measurable by observers in the interior of the black hole
horizon is constrained by the black holes's horizon area.
One can then imagine that an arbitrarily large amount of information
may be stored in the region to the future of the horizon independent of 
its surface area.

One very plausible scenario, which is supported by several 
semiclassical calculations, is that there is a bounce as the 
collapsing star nears what would be the classical singularity, 
leading to the formation of a new expanding region of spacetime
which could contain an arbitrarily large amount of information 
(measured from the point of view of internal observers.)  Given the
possibility of making a transition back to an inflationary phase this
region could resemble our universe.  

What will then happen when the horizon evaporates.  In such a case 
there is no real spacetime singularity, and there is correspondingly 
no need for an event horizon.  This means that attempts 
to construct a paradox by making small perturbations to the usual 
black hole global structure, which do not eliminate the singularity, 
are likely of no relevance to the real physical problem.  There will 
be an apparent horizon and  under evaporation it will shrink to a size at which 
quantum fluctuations of the  gravitational field will be significant.  
At this point one will have a small wormhole, linking our spacetime to 
the origin of a large inflating region.  Most of the information that 
went into the black hole will be trapped in the new region, but there 
will be no local violation of any physical principle.  This does not mean 
that there cannot be global unitary evolution in the whole spacetime, 
but only that not all measurements made in the interior of the bulk 
can be communicated to null infinity.  

Is this kind of scenario plausible?  This is one of the key questions 
as we investigate what form a holographic principle could take in a 
gravitational theory, in which only the weak and null cosmological
entropy bounds survive.

\section{The null holographic principle}

If we give up on the possibility of a strong holographic principle 
that could hold in either a gravitational or cosmological theory, we
are forced back to the next strongest possibility, which is to 
construct a form of the holographic principle which would extend the 
null form of the cosmological entropy bound proposed by Bousso.
Since that bound is only known to hold at the semiclassical level in a
fixed cosmological spacetime $({\cal M},g_{ab})$, let us ask what
form such a null holographic principle would have to take in this case.

The problem is clearly to find a collection of light sheets that 
cover the spacetime so that the evolution of matter fields may be
described in terms of them.  What is needed can be defined as follows.

\begin{itemize}

	\item{}A classical spacetime $({\cal M},g_{ab})$ has a {\it single null
	holographic structure} if there exists a one parameter (continuous or 
	discrete)  family of
	screens ${\cal S}(t)$ with a corresponding one parameter family of 
	light sheets ${\cal L}(t)$, (each possibly made by joining two 
	lightsheets of ${\cal S}(t)$), such that for any two times, $s$ and 
	$t$, the classical or quantum state of the matter on ${\cal L}(s)$
	is completely determined by that on ${\cal L}(t)$.  In the
	quantum mechanical case, this means there is a one parameter
	family of Hilbert spaces,
	${\cal H}(s)$, which satisfies the bounds
	\f
	dim {\cal H}(t) \leq e^{A[{\cal S}(t)]/4G\hbar}
	\label{dimt}
	\ff
  such that there is for each $s$ and $t$ a unitary operator
	\f
	U(s,t) \circ {\cal H}(t) = {\cal H}(s)
	\label{unitnull}
	\ff
\end{itemize}

This is the minimal requirement, if there is going to be a 
representation of the quantum dynamics of matter in the spacetime
$({\cal M},g_{ab})$ that captures the basic principles of ordinary quantum 
mechanics.  

The problem is that such a structure does not exist for generic 
spacetimes $({\cal M},g_{ab})$.  By (\ref{dimt}) and
(\ref{unitnull}) we see that all the screens in the family must have
the same area, otherwise their Hilbert spaces cannot be unitarily 
equivalent.  The problem is that in generic spacetimes the lightsheets
of any single screen will not cover the complete future or past of
any Cauchy surface.  The reason is that the lightsheets are compact, 
and of limited extent.  This is in fact the whole point of Bousso's
bound.  Consequently, given any two screens ${\cal S}(s)$
and ${\cal S}(t)$ it will almost never happen that the corresponding
light sheets ${\cal L}(s)$
and ${\cal L}(t)$ form a {\it complete pair}. By a complete 
pair\cite{FM3,FM4,FM5} is meant a pair of non-timelike surfaces such that
${\cal L}(s)$ is within the causal future of ${\cal L}(t)$ and is 
complete in that no event can be added to ${\cal L}(s)$ which is also in the 
causal future of ${\cal L}(t)$, which is acausal to ${\cal L}(s)$, 
and the same is true reversing $s$ and $t$ and past and future.  

It is only between complete pairs that one can expect to find 
deterministic evolution in either a classical or quantum theory on a 
fixed spacetime.  

In a few very special cases involving highly symmetric spacetimes, one can
find such a single null holographic 
structure\cite{raphael,raphaelreview}.  
But these are special 
cases in which the symmetry allows the lightsheets to be complete 
futures of Cauchy surfaces.  One can say that in highly symmetric 
spacetimes such as Minkowski or DeSitter spacetime complete lightsheets
can exist because by the symmetry there is so little information for 
an observer in the spacetime to measure. Once any inhomogeneity is turned on we 
expect that the light surfaces will contract to finite regions and 
any two will be very unlikely to make a complete pair. But what is 
required generically is not only  
that we have a family of light surfaces any two of which make a 
complete pair. In addition the screens of those light surfaces must 
all have the same area.  There is no reason to believe these 
conditions can be satisfied in a generic spacetime.

Can we weaken the condition?  We can if we give up the idea that there 
is a one parameter family of light surfaces, each of which has a Hilbert spaces, 
all of which are unitarily 
equivalent.  This idea  conserved the structure of ordinary 
quantum mechanics in which there is a single Hilbert space on which 
evolution is unitarily implemented.   However it is clearly ruled out.

For a generic spacetime it is clear that the lightsheets of no screen 
will be complete in the future or past of any Cauchy surface.  In this 
case if we want a description in terms of screens we must allow the 
possibility that a complete description of the system will require 
generally more than one screen, representing information available to 
different local observers in the spacetime.  This means that a complete
holographic
description of a quantum field theory in a  cosmological spacetime system will 
generically involve multiple Hilbert spaces, each of which represents
information available to observers at different screens.  Time 
evolution must then be represented in terms of maps between density 
matrices in these HIlbert spaces.  Unitary evolution will only be 
possible for pairs of such Hilbert spaces that describe causal domains 
that form complete pairs.

Is such a multiple Hilbert space description of a quantum theory in a 
cosmological spacetime possible? In fact exactly such a structure was 
proposed in \cite{FM3,FM4,FM5}, under the name of quantum
causal histories.  It arose from an independent line of thought, coming 
from attempts to take seriously the limitations on the algebra of
observables coming from the causal structure of relativistic 
cosmological theories.  As we showed in \cite{screens} this structure 
does admit a formulation of a weak holographic principle.  

\section{Is every two surface a screen?}

In some approaches to the cosmological holographic principle, screens
are two surfaces satisfying special conditions.  Such conditions
are also used to distinguish which side of a two-surface may be
a screen, for example for screens in normal regions in Bousso's
approach, only one side of a surface will in general be a screen.

It is then important 
to ask whether any such conditions may be imposed in the case
of quantum cosmology.  There seems to be a problem with each of the
possible conditions that have been offered at the semiclassical level.

\begin{itemize}
	
	\item{}As there are no asymptotic regions, and no boundaries to a
	cosmological spacetime, there are no global event horizons.  Generic
	spacetimes do not generically contain any
	single null holographic structures, which means that the information 
	measurable on any screen cannot be used to completely specify the 
	state of a classical or quantum cosmology, and a holographic 
	principle cannot be formulated in terms of any single one parameter 
	family of screens.
	
	\item{}The operator that measures the convergence of null rays at a 
	surface does commute with the operator that measures its area.  This 
	could be used in a quantum theory of gravity to distinguish the two 
	sides of a screen.  However, this does not have the same implications 
	in the quantum theory because the local positive energy conditions on the 
	energy momentum
	tensor do not hold even at the semiclassical level.  Because of this
	null rays may diverge after beginning to converge and trapped surfaces 
	cannot be distinguished by any local conditions.  Furthermore,
	the operator that measures the convergence of a null 
	ray a finite distance from the surface does not commute with its
	convergence on the surface.  This means that in a quantum theory one 
	cannot apply the tests we use in the classical theory to pick out a 
	lightsheet.  Consequently there seems to be no reason in the quantum 
	theory to choose one 
	side of a screen over another.  
	
	\item{}No condition can be imposed having to do with the volume of
	a spacelike region bounding a screen, for the volume of a region
	is measured by a quantum mechanical operator\cite{vol1,sn1}.  
	Generic states will
	be superpositions of eigenstates of the volume operator for any
	region.  One can thus not require that the side of a two-surface which
	encloses the smallest volume is a screen.
	
	\item{}As we see from several of the counterexamples, 
	there is no paradox in considering both sides of a surface to 
	be a screen, so long as one understands the entropy bound weakly, 
	so that  it applies only to information gained by making measurements of
	fields at the surface, which may or may not allow deductions to be 
	made concerning the density matrix or state to the causal past of the 
	surface.
	
\end{itemize}	

If there is no criteria which can be applied in a quantum theory
of cosmology to pick out which surfaces are screens, or to pick one
of the two sides of a two-surface to serve as a screen, then we must
conclude that every two-surface
may be a screen, and the opposite side of any screen may also be a 
screen.   In the 
quantum theory one may still make observations on a screen, but one 
will not in general be allowed to deduce anything about the extent to 
which those observations allow a complete description of the physics 
on a finite lightsheet.  Since that was the reason to prefer one 
screen over another, the conclusion is that in the quantum theory if a 
screen is a useful concept, then all two surfaces may be screens.

This conclusion will play an important role in the weak holographic
principle because it means that in a quantum theory we may use the 
properties of a screen
as a place where measurements may be made to constrain, or even 
define,its geometrical properties, rather than the reverse, which is 
what we do in the semiclassical theory.

\section{Conclusions reached so far}

To motivate the weak form of the holographic principle we summarize
the results of the argument so far. 

\begin{itemize}

	\item{}The strong entropy conjecture is apparently false, which means
	that the weak, rather than the strong version of the Bekenstein 
	bound is true.
	
	\item{}The strong cosmological entropy bound is false.
	
	\item{}The  null cosmological entropy bound
	cannot be formulated in a quantum theory of gravity
	once the gravitational degrees of freedom are turned on, at
	least in the conventional terms in which entropy is related to 
	the lack of purity of density matrices.
	
	\item{}The weak cosmological entropy bound may be satisfied
	in a quantum theory of gravity.    This is formulated as 
	a relationship between the information capacity of a screen $\cal S$,
	as measured by the dimension of the Hilbert space ${\cal H}_{\cal S}$
	which provides the smallest faithful representation of the algebra of
	observables ${\cal A}_{\cal S}$ on the screen, and its area $A[{\cal 
	S}].$
	
	\item{}From the wiggly surface problem we learn that the appropriate
	measure of the area of a screen is not the area of $\cal S$.
	Instead, the amount of information that can be
	stored on any screen, $\cal S$  is bounded by  
	the minimal area of the cross sections
	of congruences of light rays that intersect $\cal S$.  This means
	that a causal structure is required in order to make sense of
	a holographic bound in a quantum cosmological theory.
	
	\item{} The information
	coded on a screen $\cal S$ then concerns its causal past.  But it 
	then follows that in most histories there will be no single screen 
	on which a complete description of the universe may be coded, for there
	will, in the classical limit, be generally no spacelike two-surface
	such that the past of its lightsheets contains a Cauchy surface.  (We 
	see this also from the throat and inflationary examples.)
	From this it follows that a holographic description in a quantum
	cosmology must involve many screens ${\cal S}_{i}$, 
	and that the information available
	at any one screen will almost always be incomplete. 
	
	\item{}One implication of this is that the most complete description 
	of the quantum
	state available on any single ${\cal S}_i$ must be a density matrix
	$\rho_i$ on ${\cal H}_i$.  This is because there will in general
	be quantum correlations that connect  measurements made on 
	${\cal S}_i$ with
	degrees of freedom that are recorded on other surfaces
	${\cal S}_j$.

\end{itemize}

\section{The weak holographic principle}

A weak form of the holographic principle must be consistent with these
conclusions.  One possible form, which is, is that given in 
\cite{screens}.  In somewhat less technical language than that give 
there, the principle holds that

\begin{enumerate}
    \item
    A   holographic cosmological theory must based on a causal 
    history, that is,  the events in the quantum spacetime form a 
    partially ordered set under their causal relations. 
	
    \item
    Among the elements of the quantum spacetime, a set of 
    screens can be identified.  A screen $\cal S$, is a  2-sided object,
	which means that it consists of a left and right side, each of which
	has a distinct past and future, but such that the past right side is
	to the immediate past of the future left side, and visa versa.
	
    \item
    Associated to each side of the screen, labeled $L$, and $R$ are
	an algebra of observables, ${\cal A}^{L.R}_{\cal S}$ each of which
	is represented on a finite dimensional Hilbert space 
	${\cal H}_{\cal S}^{L,R}$. The observables in ${\cal A}_{L.R}$
	describe information that an 
    observer at the screen may acquire about the causal past of 
	one side of the 
    screen, by measurements of fields on that side of 
    the immediate past of the left or right side of the screen.  
	
    \item
    ${\cal H}_{\cal S}^{L}= {\cal H}_{\cal S}^{R \ \dagger}$, which
	means they have the same dimension.
	
    \item
    All observables in the theory are operators in the algebra of 
    observables ${\cal A}({\cal S})$ for some screen $\cal S$. 
	
    \item
    The area of a screen $\cal S $ is {\it defined} to be 
    \f
    A[{\cal S}]  \equiv 4G\hbar \ln Dim \left ( {\cal H}_{\cal S}
	\right )
    \label{eq:yes}
    \ff    
	 
\end{enumerate}

More discussion of this principle may be found in \cite{screens}.   
Its message is that
{\it all} observables in a quantum theory of cosmology are associated with
two-surfaces, and represent information reaching a surface from
its causal past.  Besides the logic we have followed here, there are
two sets of arguments that might be used to support this hypothesis.

\subsubsection*{Quasi-local quantities in classical general relativity}

 Even in classical general
relativity, it is well understood that diffeomorphism invariance
and the equivalence principle forbid the possibility of local
definitions of the basic dynamical quantities such as energy,
momentum and angular momentum.  These kinds of quantities can only
be defined in terms of integrals  over two dimensional surfaces
in the spacetime.  When those surfaces are taken to the boundary,
in non-cosmological spacetimes, these become the well known
asymptotic definitions of energy, momentum and angular momentum.
However, even in cosmological spacetimes where there are no boundaries
one may define what are called {\it quasi-local} 
observables\cite{roger-quasi,otherquasi}, in which
the energy, momentum and angular momentum of an arbitrary region are
defined in terms of certain integrals over its boundary. Since
Penrose's original suggestion\cite{roger-quasi} many different
proposals have been made for such quasi-local 
observables\cite{otherquasi}.

If there are to be non-trivial notions of energy, momentum and
angular momentum in a quantum theory of cosmology then, these
must be defined so that their classical limits are these quasi-local
quantities.   The simplest possibility is that the hamiltonian
in quantum gravity should itself be quasi-local, that is defined
on two dimensional surfaces, which in the classical limit
will become spacelike surface embedded in spacetime.  This implies
some form of the holographic principle, for if the Hamiltonian is
associated with surfaces there must be many hamiltonians, each
associated with a different choice of surfaces, and 
the same must be true of the algebra of
observables and the hilbert spaces on which they are represented.

\subsubsection*{Relational approaches to quantum cosmology}

Another kind of argument for the importance of surface
observables in a quantum theory of cosmology was given by 
Crane\cite{louis-preholo}, even before the holographic hypothesis
of `t Hooft and Susskind was proposed.
Crane noted the difficulties of defining a coherent measurement
theory for a quantum state ``of the whole universe'' and proposed
instead that the division of the universe into two parts-system
and observer-that is basic to Bohr and Heisenberg's measurement
theory might be relativised, so that there would be not one
quantum state of the universe, but a system of observable
algebras and hilbert spaces, one associated with every possible
splitting of the universe into two parts\cite{louis-preholo}.  

To realize this idea, Crane proposed
a categorical framework to describe the association of 
Hilbert spaces with boundaries. This was based on
positing functorial relationships between the category of cobordisms
of manifolds and the category of Hilbert spaces\cite{louis-preholo}. 
These structures are closely related to topological quantum field 
theory, as those theories can be formulated in
such categorical terms.  As topological quantum field theories
are the only class of field theories that naturally yield
finite dimensional Hilbert spaces, one may try to use them
to construct examples of holographic theories\cite{linking}.
Furthermore, as Crane pointed out, it may be
possible to extend these structures
to quantum theories of gravity because it is a fact that at both
the classical and quantum mechanical level, and for
any dimension\cite{KLR}, general relativity and
supergravity  can be understood as
deformed or constrained topological quantum field 
theories\cite{CDJ,spain,linking,pluralism,super,barretcrane,hologr,superholo}.

Crane's proposal has been an inspiration for the development of what have been
called relational\cite{carlo-rel,FM3,FM4,FM5} or pluralistic 
\cite{pluralism,morebi}
approaches to quantum cosmology.  Using the fact that general 
relativity and supergravity are constrained topological field theories,
it has been possible to realize this idea in the 
context of full formulations of quantum gravity and $\cal M$ theory 
\cite{tubes,pqtubes,mpaper}.

An even stronger version of Crane's argument was proposed recently
by Markopoulou\cite{FM3,FM4,FM5}, who noted that {\it even in classical
general relativity} the logic of propositions which can be given
truth values by observers in a closed universe is non-boolean,
because each observer can only assert the truth of falsity of
propositions about their past.  Rather than being a boolean algebra,
the algebra of propositions relevant for a classical cosmological
theory is a multivalued 
Heyting algebra\cite{FM3}. When quantized, the resulting
algebra of projection-like operators cannot be represented on a single
Hilbert space, instead, it requires a collection of Hilbert spaces,
one for every possible event at which  observations are made\cite{FM4}.
As each observer receives information from
a distinct past, the algebra of observables they can measure, and
hence the Hilbert spaces on which they represent what they
observe, must vary\footnote{Related
structures have been studied also by Isham and collaborators
\cite{chris}, who note that structures built of many Hilbert
spaces can be used to formulate the consistent histories 
proposal\cite{consistent} precisely.}.
Given the conclusions reached in the preceding sections of this
paper, this is framework is then appropriate for a formulation of 
the weak holographic principle\cite{screens}.
	
\section{Conclusions}

The conclusion of the arguments we have given here is that the  
holographic bound and holographic principle can only survive in a quantum 
theory of cosmology in their weak forms, proposed in \cite{screens}.  
While logically weaker, this form is more radical than the strong 
forms, in its implications for how a measurement theory of quantum
cosmology must be constructed. First, the weak forms require that 
causal structure exist even at the Planck scale.  This most likely 
cannot be realized in a conventional formulation of quantum  cosmology
in which the observables of the theory act on a single Hilbert space 
containing the physically allowed ``wavefunctions of the universe.''  
Instead, such a description may have to be formulated along the lines
proposed in \cite{FM3,FM4,FM5} in which there is a network of Hilbert 
spaces, each providing a representation for an algebra of observables 
accessible to a single local observer at an event or a local region of
a spacetime history. These will be related to each other by maps which
reflect the quantum causal structure.  

In such a spacetime, evolution becomes closely intertwined with the 
flow of quantum information which also defines the causal structure
at the Planck scale.  Interactions have to do with the processing of 
the information at events; as noted in \cite{FM4,FM5} a quantum spacetime
then becomes very like a quantum computer that can dynamically evolve
its circuitry.

It is then difficult to escape the conclusion that the holographic
principle, in its weak form, is telling us that nature is fundamentally
discrete.  The finiteness of the information available per unit
area of a surface is to be taken simply as an indication that 
fundamentally, geometry must turn out to reduce to counting.
Of course this conclusion has been reached independently through
other arguments coming from quantum 
gravity\cite{thooft-holo,spain,vol1,sn1,morebi}
and string theory\cite{string-dis,lenny-holo}.  But, as can be seen
most clearly from the argument of Jacobson\cite{ted-eos},
the entropy bounds and holographic principle tell us that the
description of nature in terms of classical spacetime geometry
is not only analogous to the laws of thermodynamics, it must be
exactly the thermodynamics of the fundamental discrete theory of
spacetime.

What we learn from the analysis of this paper is that in 
such a theory there is no room for the 
notion of a  bulk theory, and hence no fundamental 
role for a bulk-boundary correspondence.  There is instead a 
network of screen histories, which describe what possible observers 
might be able to observer from particular events in their spacetime.  
By averaging over histories  a bulk description may emerge
at the semiclassical level, but only as an approximation in which the
past of a particular observer can be described to first order in a 
perturbation expansion in terms of a particular
fixed classical history.  Thus the proper role of a bulk-to-boundary
map may be to serve as a correspondence principle to 
constrain the classical limit of a background independent quantum 
theory of gravity.

To put it most simply: the holographic principle is not about 
a relationship between two sets of concepts, bulk and screen
and geometry and information flow.  It is the statement that the
former reduce entirely to the latter in exactly the same sense that
thermodynamic quantities reduce to atomic physics. The familiar picture of
bulk spacetimes with fields and geometry must emerge in the
semiclassical limit, but these concepts can play no role in the 
fundamental theory.

Can this picture be used to construct a realistic quantum theory of
gravity which addresses also the other problems in the subject?  As
mentioned in \cite{screens} an example of such a theory is provided by 
a class of background independent membrane theories proposed in 
\cite{tubes}. These extend the formalism of loop quantum gravity
in a way as to provide a possible background independent form of 
string theory\cite{pqtubes,mpaper}.  So the answer is a very 
provisional, yes. Much work remains to be done, but the moral is that 
the holographic principle, in at least its weak form, is likely to 
feature significantly in 
both the mathematical language and the measurement theory of the future
background independent quantum theory of gravity.

\section*{ACKNOWELDGEMENTS}

I would like to thank
Tom Banks, John Barrett, Michael Douglas, 
Willy Fischler, David Gross, Sameer Gupta, 
Chris Isham, Jerzy Lewandowski, Yi Ling, Renate Loll, 
Amanda Peet, Joe Polchinski, Roger Penrose, Carlo Rovelli, 
Andrew Strominger,
Leonard Susskind, Edward Witten and especially Ted
Jacobson and Mike Reisenberger
for discussions  on these issues.  I would also like to thank
Raphael Bousso for very helpful correspondence and discussions.
This paper owes a great deal to conversations over many years with 
Louis Crane and Fotini Markopoulou,
who proposed several of the ideas which are
developed here.  Opportunities to present this argument to the 
topos discussion group at Imperial College and a seminar at the 
faculty for the philosophy of science at Oxford were very helpful in 
sorting out its fine points.
Finally, I am grateful for the hospitality of 
ITP, Santa Barbara, where this work was begun and of
the theoretical physics group at Imperial College, where it was 
finished. 
This work was supported by the NSF through grant
PHY95-14240 and a gift from the Jesse 
Phillips Foundation.

\end{document}